\newcommand{\beq}{\begin{equation}}
\newcommand{\eeq}{\end{equation}}
\documentstyle[aps,preprint,psfig]{revtex}

\begin{document}
\draft

\title{Stability analysis of static solutions in a Josephson
 junction}

\author{ Jean-Guy Caputo$\ ^{a,b}$, Nikos Flytzanis$\ ^{b}$\\
Yuri Gaididei $\ ^{b,c}$, Nikos Stefanakis $\ ^{b,e}$ and Emmanuel
Vavalis $\ ^{d}$\\ \ \\ {\normalsize $\ ^a$ Laboratoire de
Math\'ematiques, INSA and URA CNRS 1378,}\\ {\normalsize BP8,
76131 Mont-Saint-Aignan Cedex, France.}\\ {\normalsize $\ ^b$
Physics Department, University of Crete,71003 Heraklion,
Greece.}\\ {\normalsize $\ ^c$ Institute for Theoretical Physics,
252143 Kiev, Ukraine.}\\ {\normalsize $\ ^d$ Mathematics
Department, University of Crete, 71409 Heraklion, Greece.} \\
{\normalsize and IACM, FORTH, Heraklion, Greece.}\\ {\normalsize
$\ ^e$ IESL, FORTH, Heraklion, Greece.}}

\date{\today}

\maketitle
\begin{abstract}
We present all the possible solutions of a Josephson junction with
bias current and magnetic field with both inline and overlap
geometry, and examine their stability. We follow the bifurcation
of new solutions as we increase the junction length. The
analytical results in terms of elliptic functions in the case
of inline geometry, are in agreement with the numerical
calculations and explain the strong hysteretic phenomena typically
seen in the calculation of the maximum tunneling current. This
suggests a different experimental approach based on the use,
instead of the external magnetic field the modulus of the elliptic
function or the related quantity the total magnetic flux to avoid
hysteretic behavior and unfold the overlapping $I_{max}(H)$
curves.
\end{abstract}

\tighten

\section{Introduction}
The static properties of a narrow Josephson junction are well
characterized by the static sine-Gordon equation \cite{Barone}.
They are experimentally measured by the maximum tunneling current
$I_{max}$ as a function of the external field $H$. This is an
important and useful measurement since it is required not only for
the characterization of the junction properties but also to tailor
a device with the desired maximum current. In contrast to its
simple form, the static sine--Gordon differential equation problem
poses several mathematical and computational challenges. The
complete analysis of all of its solutions is hard due to various
interesting properties (nonlinearity, non definiteness,
periodicity, boundary conditions of Newmann type) inherent to the
sine-Gordon problem and the determination of  $I_{max}$ either
theoretically, numerically or experimentally is difficult.

Several studies analyzing the dynamic and static stability of
fluxons in the sine-Gordon equation have appeared in the
literature in the past two decades with
\cite{cr73,rwd76,ffs91,bfmp94} being the most representative of
them. All these studies combine theoretical and numerical analysis
and they mainly address the case where there is no external
current or magnetic field applied on the junction. None of the
above studies is comprehensive enough as far as exploiting all
solutions and studying the affect of all physical and geometrical
parameters of the problem.

The main objective of this study is fourfold.
\begin{itemize}
  \item To
analytically express all static solutions of 1-dimensional narrow
Josephson junctions in a way that will allow us to examine their
stability properties and their evolution with respect to the size
of the junction, and the applied magnetic field and current.
  \item To explain the hysteretic behavior and if possible to find
  the important physical parameters that unravel the hysterisis.
  \item To build a numerical simulation framework that will allow
  us to verify some of our theoretical results and show that they apply to more
  complicated Josephson junction configurations.
  \item To propose an experimental procedure that will enable us to examine the properties of superconducting devices in a more
accurate way. This approach is particularly useful for the
analysis of devices that deviate from the standard mathematical
model currently used i.e. junctions with impurities,
inhomogeneities, ...
\end{itemize}

We should mark here that, at the present, a complete theoretical
analysis of such devices is not feasible while numerical
simulations, based on state-of-the-art software packages
\cite{IJMPC95,pltmg}, usually fail to capture all the important
features. This is mainly due to the difficulty to track the
continuation branches and to deal with the bifurcation points
involved.  It is worth to mark here that the effects of the above
mentioned difficulties are clearly seen, even in a relatively
simple case, by the fact that it was only very recently
\cite{bfmp94} realized that the critical value of the bias current
corresponds not to a termination point, as conjectured for many
years, but to a turning point in the bifurcation diagram.

In addition to the above, the experimental analysis and the
computer simulation and analysis of superconducting devices
modeled by the sine-Gordon equation require an initial guess that
is reasonably close to the desired solution. The selection of such
initial guesses significantly affects the effectiveness of the
various continuation methods needed to determine $I_{max}$. These
guesses will be extensions of the various 1-D solutions obtained
here. Therefore the results of the present study are expected to
be fully utilized in effectively analyzing two dimensional window
Josephson junctions.

\begin{figure}
  \centerline{
     \psfig{figure=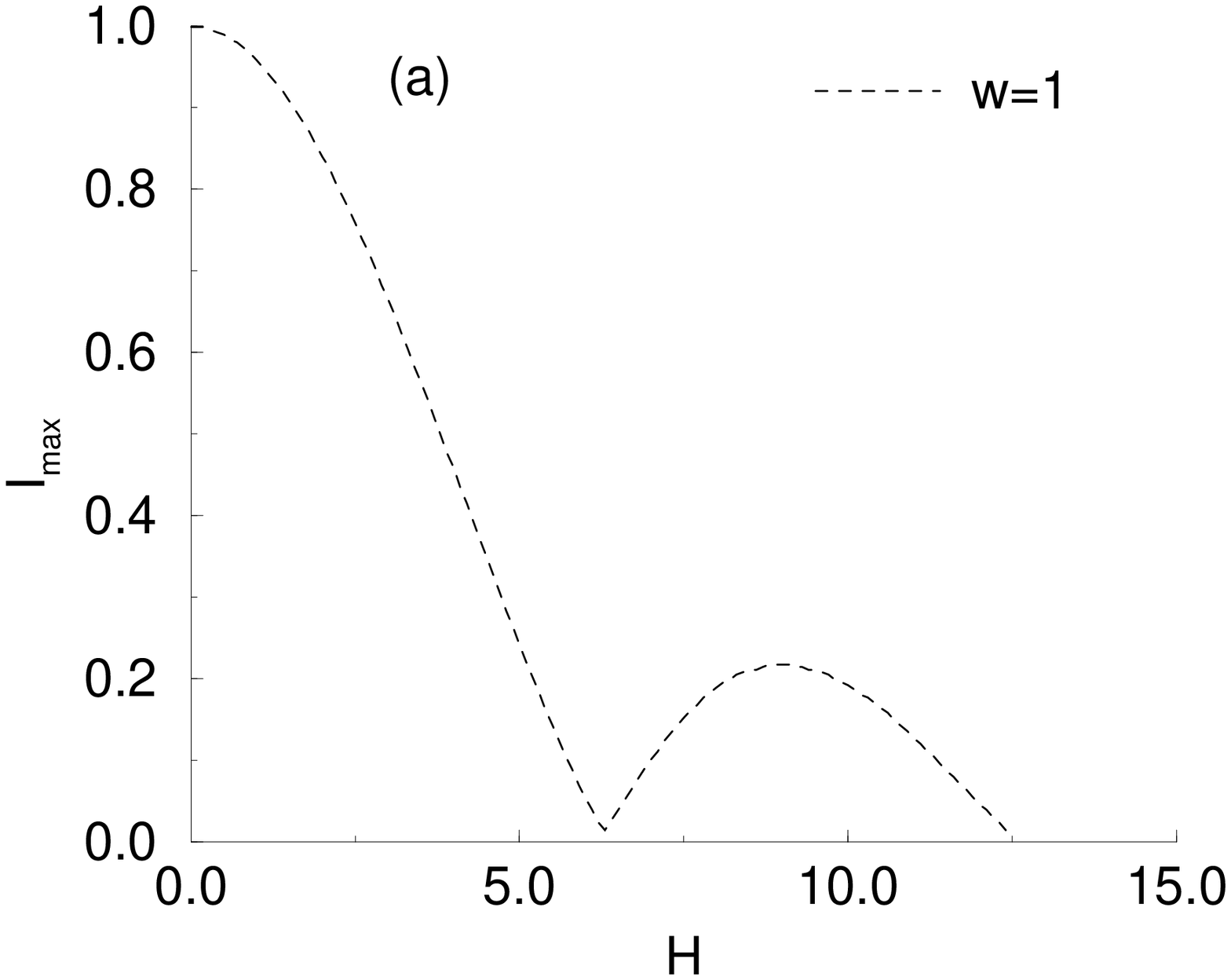,height=2.85in,width=3.1in}
             }
  \centerline{
     \psfig{figure=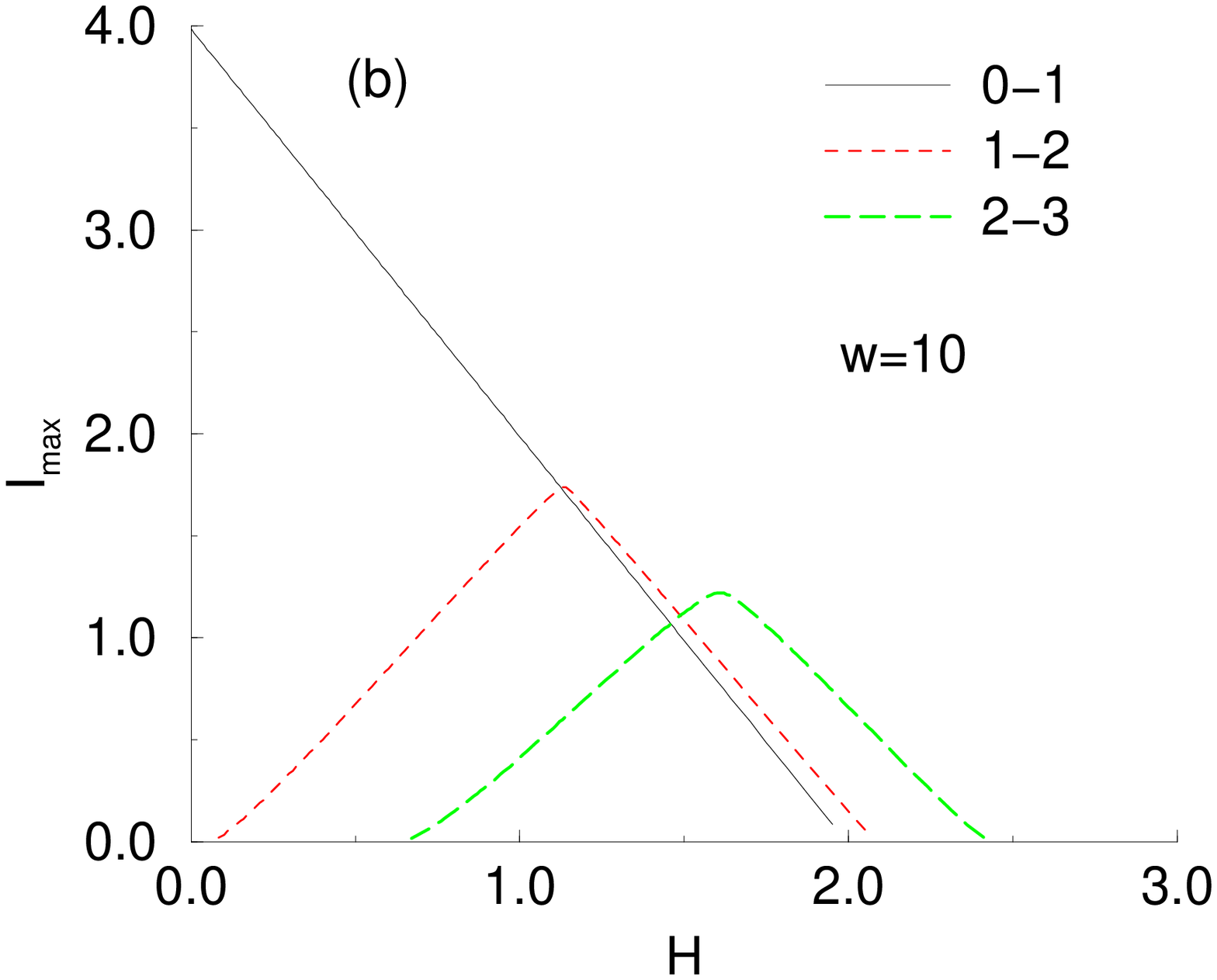,height=2.85in,width=3.1in}
             }
  \caption{ Plot of the maximum tunneling current as a function of
      applied magnetic field for a short junction with: (a) $w=1.0$ and 
(b)
      $w=10$.}
  \label{fig:1}
\end{figure}

The behavior of a Josephson junction is determined by the
Josephson characteristic length $\lambda_J$ which depends both on
material and geometry properties. For a short junction of length
$l\ll \lambda_J$ the $I_{max}$ vs $H$ pattern (shown in
Fig.~\ref{fig:1}a for normalized length
$w=\frac{l}{\lambda_J}=1.0$) presents the usual Fraunhoffer
pattern $$I_{max}=\frac{sin\pi \frac{\Phi}{\Phi_0}}{\pi
 \frac{\Phi}{\Phi_0}},$$
where $\Phi=Hld$ is the applied flux,
$\Phi_0=\frac{\hbar}{2e}$ the flux quantum and $d$ the magnetic
thickness \cite{Barone}.  Each of the lobes in the diagram can be
labeled by the pair of integers $(n,n+1)$ where at one end the
 magnetic field corresponds to exactly $n$ fluxons (i.e. flux= $n
\Phi_0$) and at the other end $n+1$ fluxons. For the case of a long
junction the problem was solved by Owen and Scalapino \cite{Owen}, and
there, the different lobes overlap (as shown in Fig.~\ref{fig:1}b for $w=10$).

It should be remarked  that the sine-Gordon
 equation, due to its nonlinearity and periodicity with several
equilibrium points, has a multiplicity of solutions as shown by the
overlapping lobes  (in Fig.~\ref{fig:1}b) and the existence of several unstable
branches which play an important role in the hysteretic behavior as
you vary the external magnetic field. The unstable branches are of
interest too because one can stabilize them, by introducing small
defects, and therefore they should lead to observable maximum current.
As we will see
later in the discussion a defect can modify significantly the relative
amplitude of the different lobes in the $I_{max} ~~{\rm vs} ~~H $
curve.
The unstable branches can be partially traced experimentally if we
perform a quasistatic scanning of the magnetic field.

The rest of the paper is organized as following. In section 2 we
present the mathematical problem, give the explicit analytical
solutions using elliptic functions and sketch the stability
analysis. In section 3 we study  the solutions in the particular
case of zero magnetic field $H=0$, and in section 4 the analytic
solutions for zero current $I=0$. We study both theoretically and
numerically  their stability in section 5 and we calculate the
maximum tunneling current. In section 6 we briefly propose an
experimental procedure which utilizes our numerical procedure. We
summarize our results in the last section.
\section{Inline Geometry and Stability of Solutions} 
The case of
inline geometry is a one dimensional problem, even for a wide
junction, and one can obtain analytic solutions \cite{Owen}.
Furthermore one can  easily check their stability by using linear
perturbations. The particular cases  of zero external magnetic field
and inline bias current also reduce the overlap boundary conditions
to inline ones.
In the 1-dimensional case we have to solve the following problem
\beq - \frac{d^2
 {\Phi}(x,t)}{dt^2} + \frac{d^2 { \Phi}(x,t)}{dx^2} = \sin {
\Phi}(x,t) ,~~~\label{eq01} \eeq with the inline boundary condition
\beq \frac{d { \Phi}}{dx}\left|_{x=\pm {w}/2}\right. =\pm \frac{{
I}}{2}+H \equiv \gamma_{\pm},  ~~~~\label{eq02} \eeq
where $w$ is the normalized junction length and ${ I}$ is the
current line density.

Eq. (\ref{eq01}) has a static solution $\Phi_0(x)$, implicitly
expressed using elliptic functions \cite{AS} as
\beq \sin \Phi_0(x) =- 2 \sqrt{m
}~sn\left(x+x_0|m\right) ~dn\left(x+x_0|m\right), ~~~~\label{eq03}
\eeq
\beq \cos \Phi_0(x) = 2 {m }~sn^2\left(x+x_0|m\right) -1, ~~~~~
\label{eq04} \eeq
\beq \frac{d \Phi_0}{dx} = 2 \sqrt{m}~
cn\left(x+x_0|m\right),  ~~~\label{eq05} \eeq
where the modulus $m$
determines the period of the $cn$ elliptic
 function (equal to $4K(m)$) and the arbitrary constant $x_0$ the
 phase at the center of the junction. They are determined by the
boundary conditions (\ref{eq02}).  Introducing (\ref{eq05}) into
(\ref{eq02}) we get 
\beq  ~~~~~~~~~~~~2 \sqrt{m}~
cn(x_0+\frac{w}{2}|m)  =  \gamma_{+}, ~~~\label{bca}
\eeq  
\beq~~~~~~~~~~~~2 \sqrt{m}~
cn(x_0-\frac{w}{2}|m)  =  \gamma_{-}.~~~\label{bcb}
\eeq

A useful quantity to classify the solutions is the fluxon content or
the magnetic flux in units of the quantum of flux, defined by
$$N_f=\frac{1}{2\pi}\left\{\Phi(\frac{w}{2})-\Phi(-\frac{w}{2})\right\}.
$$
At specific values of $H$, $N_f$ takes integer values, so that the
flux is that of an integer number of fluxons.

To check the stability of the static solution
(\ref{eq03})-(\ref{eq05}) we consider small perturbations on the
static solution $\Phi_0(x)$ in the form \beq \Phi(x,t) = \Phi_0 (x) +
U (x,t)~~~~~~~~    \label{eq07} \eeq and linearize Eq. (\ref{eq01})
with respect to $U(x,t)$, to obtain \beq -U_{tt} + U_{xx} = \cos
\Phi_0 (x) \cdot U.~~~~~~~~\label{eq08} \eeq There is no loss of
generality if we consider specific perturbations in the form \beq
U(x,t) = X(x) e^{st}.~~~~~~\label{eq09}  \eeq  This way we obtain
from Eq. (\ref{eq08}) the eigenvalue equation
\beq -X'' + \cos \Phi_0(x)
\cdot X = \lambda \cdot X,    \label{eq10}  \eeq under the boundary
conditions \beq X'|_{x=\pm w/2} = 0.~~~~~~~\label{eq11} \eeq
\noindent
In Eq. (\ref{eq10}) $\lambda= -s^2$.  It is seen from Eq. (\ref{eq09})
that if the eigenvalue problem (\ref{eq10})-(\ref{eq11}) has a
negative eigenvalue the static solution $\Phi_0 (x)$ is unstable. If
all the eigenvalues are positive $\Phi_0(x) $ is stable, while the
case $\lambda =0$ corresponds to neutral stability and defines the
boundary of stability.  In the following we consider the two special
cases $\gamma_+=\gamma_- ~(I=0)$  and $\gamma_+=-\gamma_-~ (H=0) $
separately, since the associated boundary conditions are easy to handle
and the stability analysis is simplified.

\begin{figure}
  \centerline{
     \psfig{figure=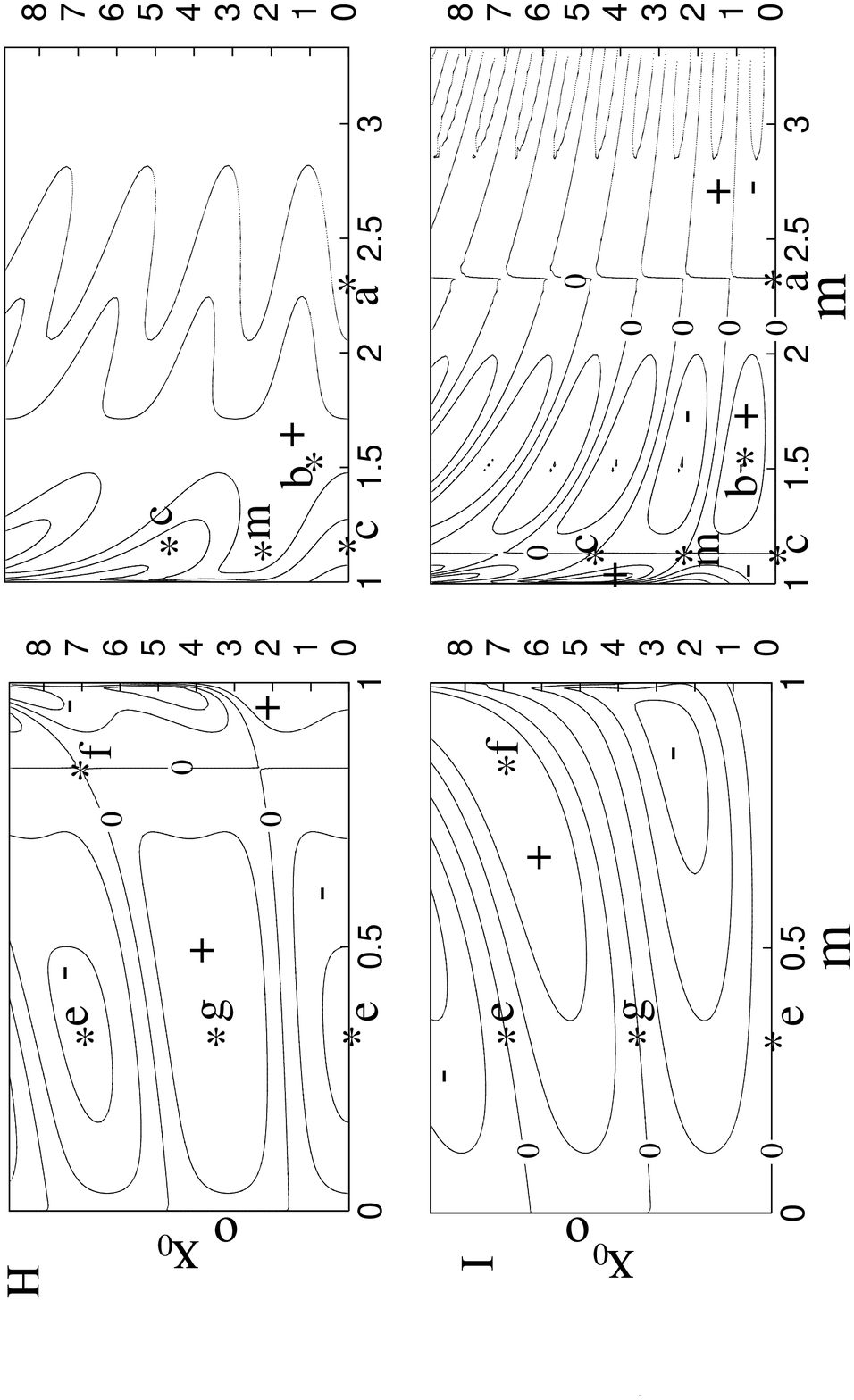,width=5.2in,angle=-90}
             }
  \caption{Constant $H$ (a) and $I$ (b) contours in the ($m$, $x_0$)
plane. The curves with the symbol $O$ are for $H=0$ in (a) and $I=0$
in (b). The signs $"+"$ or $"-"$ give the sign of $I$ and $H$. The
 letter symbols signify the same points in the $I$ and $H$ contours.}
  \label{fig:2}
\end{figure}

To get a feeling concerning the possible solutions, we plot from the boundary
conditions (\ref{eq02}) and (\ref{bca},\ref{bcb}) the constant $H$ and $I$ contours in
the plane of the $m$ and $x_0$ parameters in Fig.~\ref{fig:2}. We give different
plots for $m<1$ and $m>1$. The  lines labeled with "$0$" correspond to
$H=0$ (or $I=0$ for $I$ contours) and in both cases  their network
encloses areas with a single maximum (denoted by "$+$") or minimum
("$-$") inside.  Notice that there are two types of curves on which
$H=0$ (or $I=0$) as summarized in Table ~\ref{table1}.

The curved  lines in Fig.~\ref{fig:2}
correspond to the solutions of the first line in the table which we
call fixed $x_0$ solutions, in the sense that the shift is a fixed
fraction of the period. From (\ref{bca},\ref{bcb}) we see that the physical quantities
$I$ and $H$ are periodic functions of $x_0$ with  period $4K(m)$.
There are also solutions that have a fixed $m=m^{\star}$ and arbitrary
$x_0$, and correspond to the vertical  lines in Fig.~\ref{fig:2} (remark that
contour fitting can be distorted when two equicurrent lines cross).
Similar results hold for the constant $I$ contours.  For $m>1$ at the
vertical $I=0$ curves we have an integer number of fluxons in the
junction. Thus on the lines through $c$ ($a$) we have $N_f=1$ (2)
correspondingly. In the current contours the points $f$ and $b$
correspond to peaks in the $I_{max}$ (see section 4.4). In the
following we will focus our attention on the solutions where either
$H$ or $I$ .

\section{ No magnetic field case  (H=0)}

In the absence of external magnetic field we have
$\gamma_+=-\gamma_-=\frac{I}{2}$ and  Eqs. (\ref{bca},\ref{bcb}) reduce to
\beq ~~~2\sqrt{m} cn \left(x_0+\frac{w}{2}
|m\right)  =  \frac{I}{2},~~~\label{bcH=0a}
\eeq
\beq
\sqrt{m} cn \left(x_0-\frac{w}{2} |m\right) 
=  -\frac{I}{2},~~~\label{bcH=0b}
\eeq
which determine the parameters $m
$ and $x_0$ that characterize the periodicity and phase shift of the
static solutions.
There are two different classes of solutions due to the antisymmetry
of the boundary conditions, which  can be satisfied for
different $I$ either by fixing $x_0$ or $m$.

\subsection{ Fixed $x_0$ solutions}

It is seen from Eqs. (\ref{bcH=0a},\ref{bcH=0b}) and the symmetry of the
elliptic functions that  for positive $I$ one choice for $x_0$
in (\ref{bcH=0a},\ref{bcH=0b}) is $x_0=-K(m)$ ($K(m)$ is the elliptic integral of the first
kind),  so that $x_0$ is fixed to $\frac{1}{4}$ of the period of the
elliptic function. It is in that sense that we call them fixed
$x_0$ solutions. Strictly $x_0$ is not a constant independent of $m$
since $K(m)$ is a function of $m$. Thus we have (see \cite{AS}) 
\beq
\frac{I}{2}=2\sqrt{m(1-m)}~ \frac{sn (\frac{w}{2}|m)}{dn
(\frac{w}{2}|m)} ,~~~ 0\le m <1,\,\,\,~~~\label{eq13} 
\eeq 
\beq \cos
\Phi_0(x)=2~m~ \frac{cn^2 (x|m)}{dn^2(x|m)}-1 ,~~~ 0\le m <1,\,\,\,
~~~\label{eq14} 
\eeq 
\beq \sin \Phi_0(x)=2~\sqrt{m~(1-m)} \frac{cn
(x|m)}{dn^2(x|m)} ,~~~ 0\le m <1.\,\,\, \nonumber 
\eeq 
Another possibility is $x_0=K(m)$ ($x_0$ is shifted by half of a
period) for the case that $2K(m) <\frac{w}{2}<4K(m)$, or more
generally when $sn (\frac{w}{2}|m)<0$, since we are limiting ourselves
to $I>0$.  This means that every time $w$ increases by $2\pi$ we
introduce two extra solutions. To put it in other words the function
$I(m)$, in (\ref{eq13}),
is highly oscillating for large $w$. We do not need to
consider the case that $m>1$ since in that case we cannot satisfy  the
antisymmetric boundary conditions for the external current.

\begin{figure}
  \centerline{
     \psfig{figure=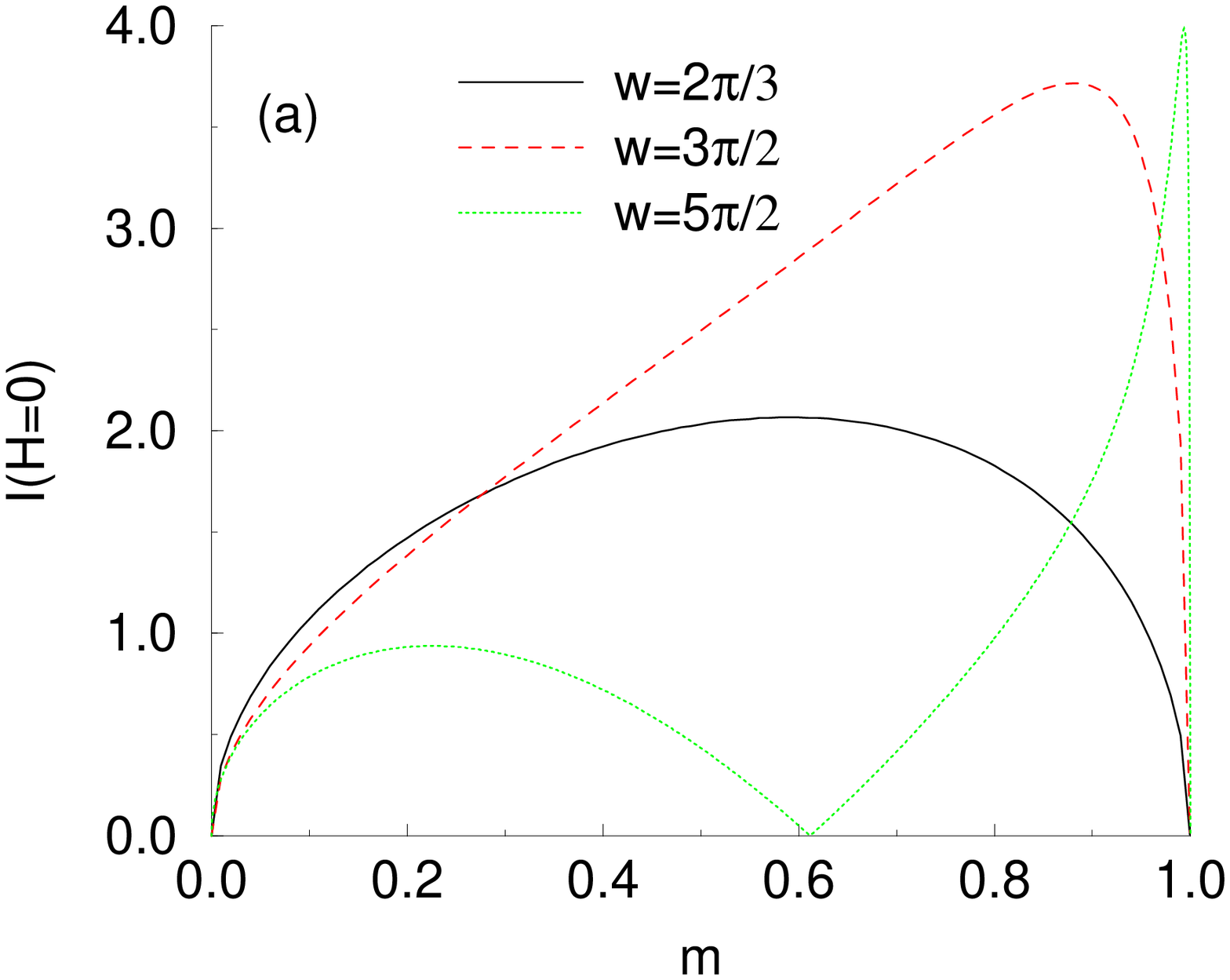,height=2.85in,width=3.1in}
             }
  \centerline{
     \psfig{figure=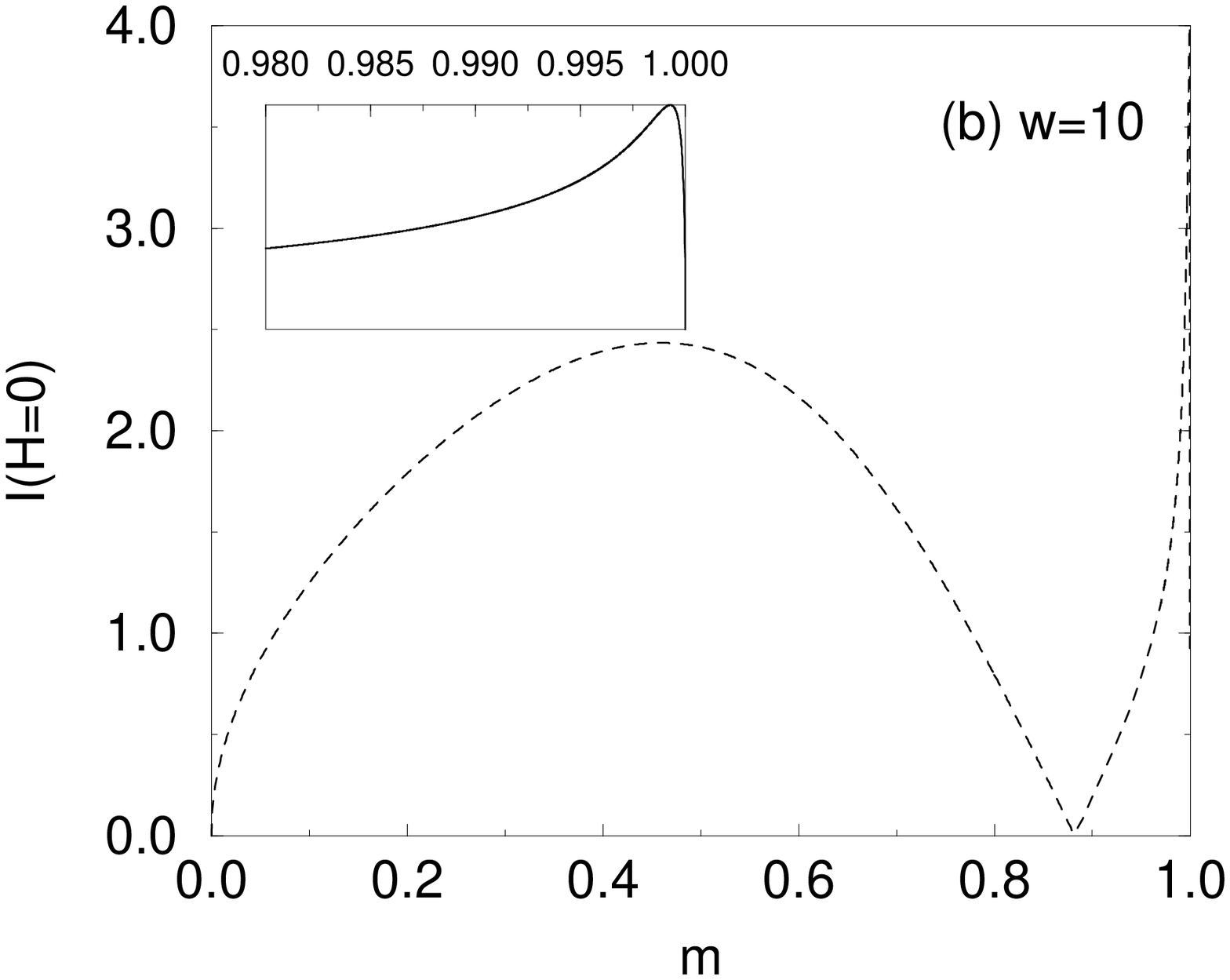,height=2.85in,width=3.1in}
             }
  \caption{(a) Plot of $I(m)$ (using Eq.  (13)) with $x_0=-K(m)$ or
 $K(m)$ and three values of $w$:  (i) $w=2\pi/3$, (ii) $w=3\pi/2$ and
 (iii) $w=5\pi/2$. (b)  Same as (a)  with $w=10$ (dashed line).  The
 right lobe in (b) is also shown in an expanded scale for $m$ (solid
 line). The small $m$ scale is shown at the top of the figure.}
  \label{fig:3}
\end{figure}

In Fig.~\ref{fig:3}a we present three plots of ${I}$ vs $m$ for
$w=\frac{2\pi}{3},\,\,\, \frac{3 \pi}{2},\,\,\, \frac{5\pi}{2}$.  We
see that for small $w<2\pi$ there is, as expected,
only one lobe for
$x_0=-K(m)$, and as we will see later only the part to the right of
the maximum will correspond to stable solutions, while the peak
corresponds to the maximum current for zero magnetic field. For
$w=\frac{5\pi}{2}$ we have an extra lobe, with the left one at
$x_0=K(m)$ and the right at $x_0=-K(m)$. For $w=10$ (dashed line
in Fig.~\ref{fig:3}b) the right lobe has a maximum within $10^{-3}$ of $m=1$ and
corresponds to $x_0=K(m)$, while the left to $x_0=-K(m)$.
The jump in path along two different curves is necessary because of the
restriction of positive current $I$. Of course the curve is symmetric
about the $I=0$ line.  The right lobe for $w=10$ is shown in the inset
of Fig.~\ref{fig:3}b in expanded form (solid line) with a different scale for
$m$.    The part that is of experimental interest is the last lobe
near $m=1$ to the right of the maximum. The two extremal $m$ values
correspond to trivial solutions $\Phi_0(x)=\pi~ (m=0)$ and
$\Phi_0(x)=0 ~(m=1)$, the first of which is clearly unstable (pendulum
analogy) and the second is stable. For currents above zero at $w=10$
there are four possible solutions for a given $I<I^{\star}$, where
$I^{\star}$ is the maximum of the lowest lobe.

Because the last lobe for large $w$ is very steep it is useful to give
some analytic formulas valid near $m=1$ and for the maximum point.
By using asymptotic formulas and assuming that $m_1\equiv 1-m\approx
\epsilon^2$, where $\epsilon $ is a small parameter, we obtain the
value of $m_1$ where $I$ is a maximum as $$
m_1^{max}=\frac{4}{\sinh^2\frac{w}{2}}. $$ The result is consistent
with our scaling assumption with $$\frac{1}{\sinh\frac{w}{2}}\sin
\epsilon. $$ Thus care is required when simplifying the analytic
formulas in  \cite{AS} (see p. 574). The corresponding maximum
current is
\begin{equation}
I= 4-\frac{8}{\sinh^2\frac{w}{2}}, \label{iw}
\end{equation}
so that for large junction length  $w$ it approaches exponentially
the infinite length limit.  To the right of the maximum the relation
between $I(m_1)$ is $$ I= 4- m_1^{3/2} \frac{\tanh \frac{w}{2}}{{\rm
sech} \frac{w}{2}}.  $$

\begin{figure}
  \centerline{
     \psfig{figure=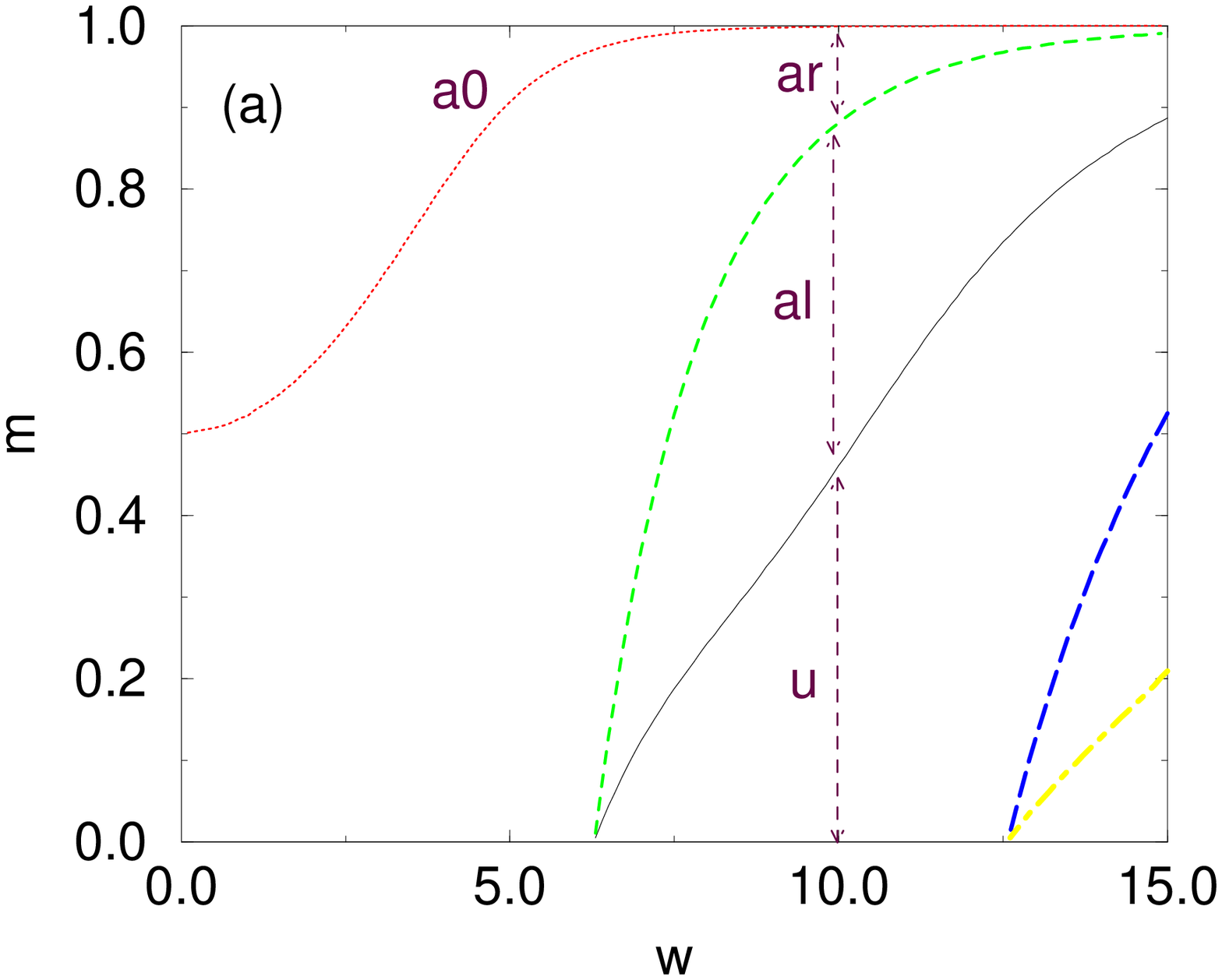,height=2.85in,width=3.1in}
             }
  \centerline{
     \psfig{figure=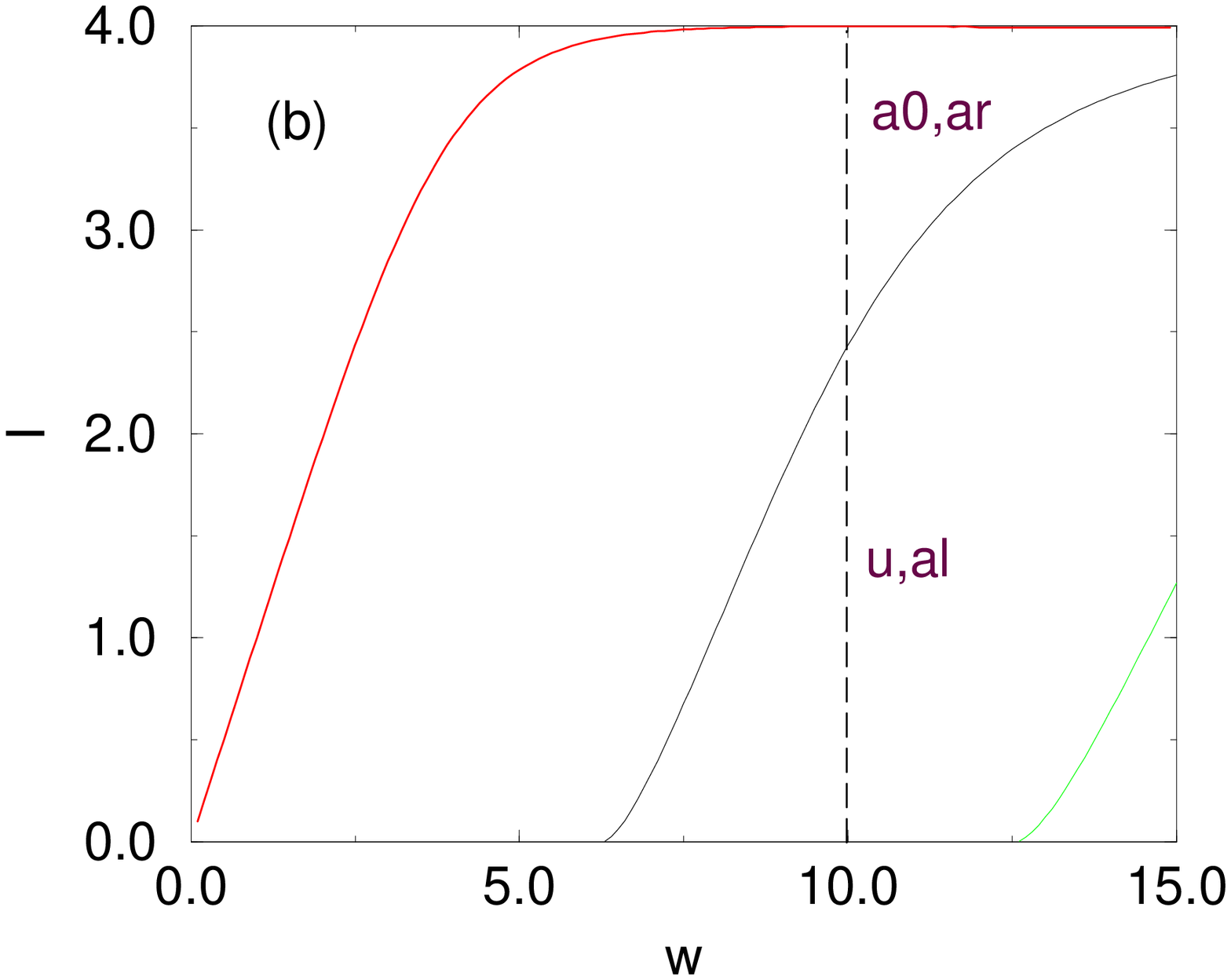,height=2.85in,width=3.1in}
             }
  \caption{(a) A diagram of the different solutions in the space of 
parameter
 $w$ and $m$ with $x_0=\pm K(m)$ for the solutions in equations
(13)-(15). (b) The same as (a) but presented in the space of $w$ and
$I$ (instead of $m$).}
  \label{fig:4}
\end{figure}

From the previous discussion we see that as we increase the junction
length $w$ we obtain more solutions. In Fig.~\ref{fig:4}a we give as a function
of $w$  the range of $m$ values  for each type of solution and the
separating lines.  These values were determined by solving 
(\ref{bca},\ref{bcb})
numerically. We remark that consecutive pairs of regions of  solutions
correspond to different $x_0$ (i.e.  different lobes of Fig.~\ref{fig:3}).  We
see that when $w$ increases by $2\pi$ a new pair of solutions is
introduced. Thus near $w=10$ we have four solutions labeled $u$, $al$,
$ar$ and $a_0$, the first two corresponding to $x_0=K(m)$ and the last
two to $x_0=-K(m)$. For $w \rightarrow \infty $ there is an infinite
number of solutions and many of the dividing lines coalesce at $m=1$.
The stability is checked by looking at the eigenvalues of the
linearized problem in (\ref{eq10}). We see that already for $w=10$ only a
small range near $m=1$ gives stable solutions, while for $w=14$ it is
of the order $10^{-7}$, which is extremely small and not visible on
the scale of the plot.
In Fig.~\ref{fig:4}b we give the same information but in a diagram of current
vs $w$. The lines correspond to the maximum current for each lobe
and below each line there are two solutions. Thus the solutions $a_0$
and $ar$ have the same maximum current (starting from zero current)
and correspond to the right lobe of Fig.~\ref{fig:3}b, while $u$ and $al$ to the
left one.
In order to
make sure that we get all solutions we scan over $m$ (with a uniform
and fine grid), and the current is obtained from (\ref{eq13}).
As expected  the maximum current for large $w$ is accurately estimated
by (\ref{iw})  down to $w=4$, while for small $w$ it varies linearly.

\subsection{ Fixed $m$ solutions}
\begin{figure}
  \centerline{
     \psfig{figure=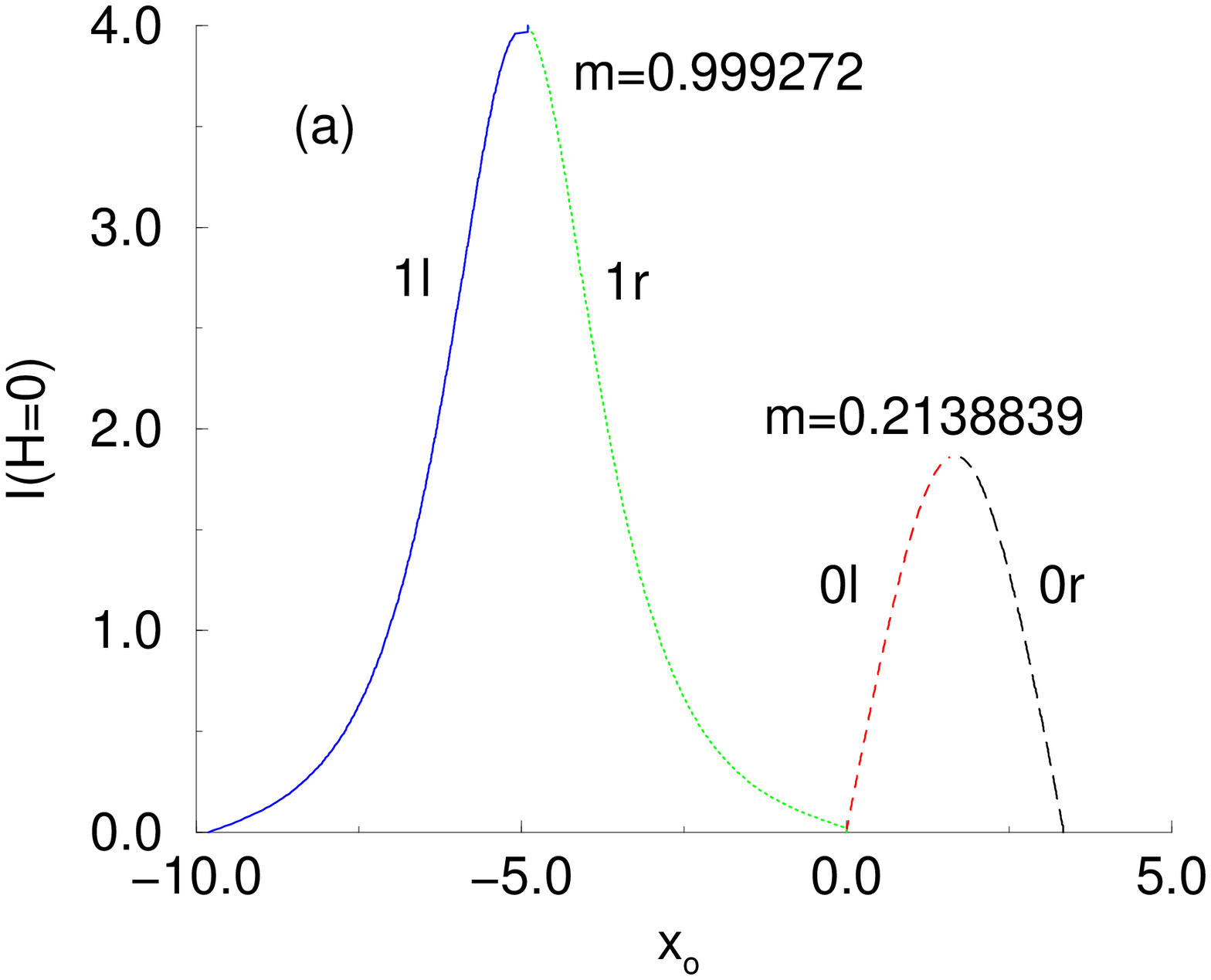,height=2.85in,width=3.1in}
             }
  \centerline{
     \psfig{figure=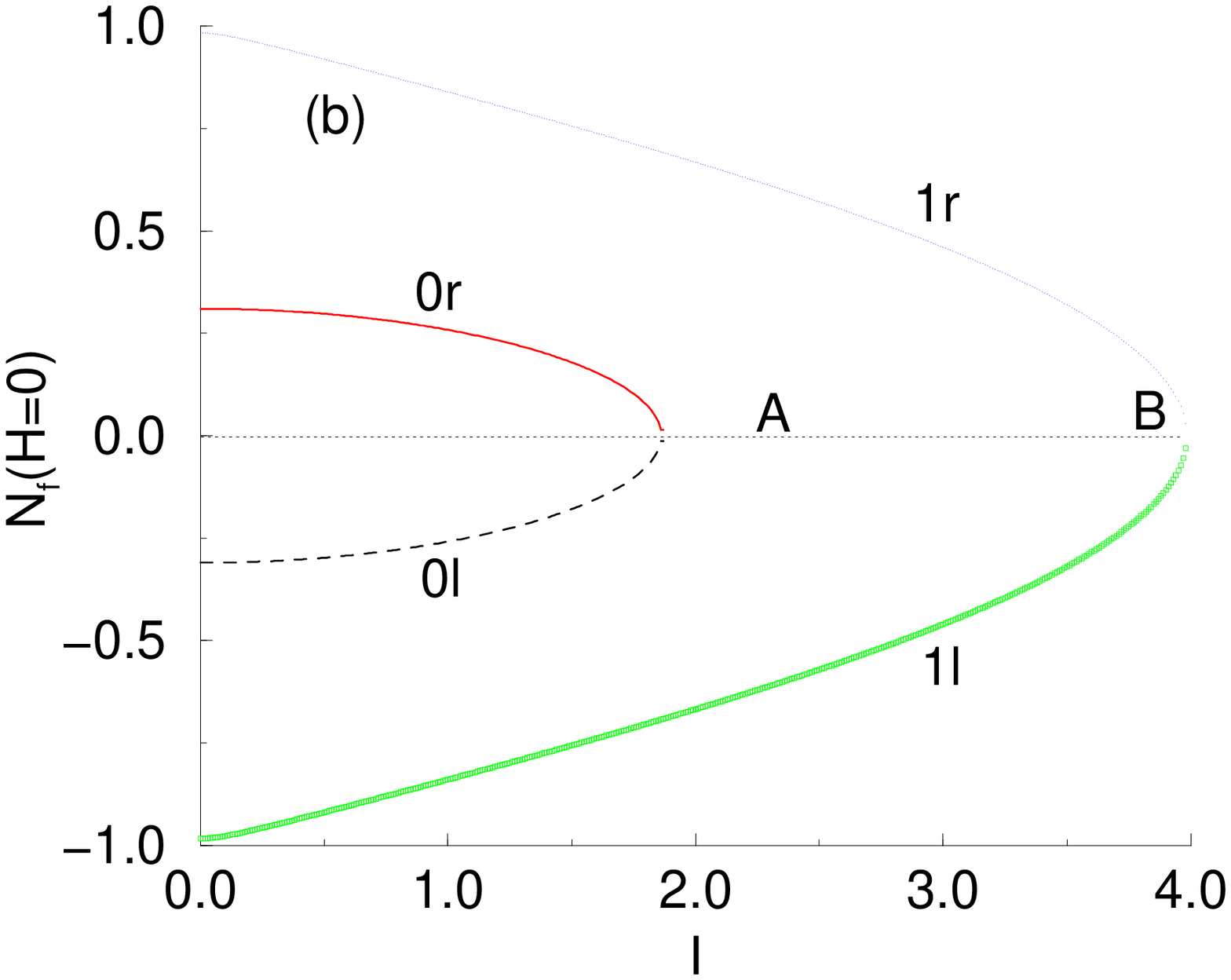,height=2.85in,width=3.1in}
             }
  \caption{(a) Plot of $I(x_o)$  for $H=0$,
       $w=10$ and fixed $m$ with (i) $m^*=0.999272$ (curves $1l$ and 
$1r$)
       from $w=2K(m^*)$, (ii) $m^*=0.213839$ (curves $0l$ and $0r$) from
        $w=6K(m^*)$.  (b) Fluxon content $N_f$ as a 
function
        of the bias current $I$ for the different solutions with fixed 
$m$:
        (i) $m$=0.999272 with one half period of the elliptic function
        ($1r$ and $1l$). (ii) $m$=0.213839 with three half periods of 
the
        elliptic function ($0l$ and $0r$).}
  \label{fig:5}
\end{figure}

Another possibility exists if $w>\pi$, so that we can fit in the
length exactly an odd number of half periods. This
automatically satisfies
the antisymmetric boundary conditions due to the
current. Then, there exists a fixed (sometimes more than one depending
on the length) $m=m^*$ for which $w=2K(m^*)$. In fact every time $w$
increases by $2\pi$ there is an extra solution arising. Thus for $\pi
(2n+1) <w<\pi(2n+3)$, we have solutions at $w= 2K(m),\cdots ,
 2(2n+1)K(m)$ with $n$-different values of $m^*$.  By shifting
$x_0$ we can obtain a range of possible currents $I$ while always
satisfying the  boundary conditions
at zero magnetic field.  In Fig.~\ref{fig:5}a we plot the
  current as a function of $x_0$, for $w=10$, where we expect two
solutions of this type. The corresponding values of $m^*$ are
$m^*=0.999272$ (from $w=2K(m^*)$, see curves $1l $ and $1r$ in
Fig.~\ref{fig:5}a) and $m^*=0.213839$ ($w=6K(m^*)$, see curves $0l$, $0r$
in Fig.~\ref{fig:5}a), with the maximum currents being $I_0\approx 4$ (at
$x_0=-\frac{w}{2}$) and $I_1 =1.8$ (at $x_0=\frac{w}{6}$)
correspondingly. These also coincide with the maximum currents
obtained by fixing $x_0= -K(m),\,\, K(m)$ and varying $m$. It should
be remarked, though that they correspond to different solutions as we
increase the current ($I<I_{max}$). This can be seen by the different
fluxon content of these solutions in Fig.~\ref{fig:5}b.  At the maximum current
value though they coincide.  In comparison, the solutions in the
previous subsection correspond to zero fluxon content $N_f=0$.

\subsection{Solutions for $w=10$}

\begin{figure}
  \centerline{
     \psfig{figure=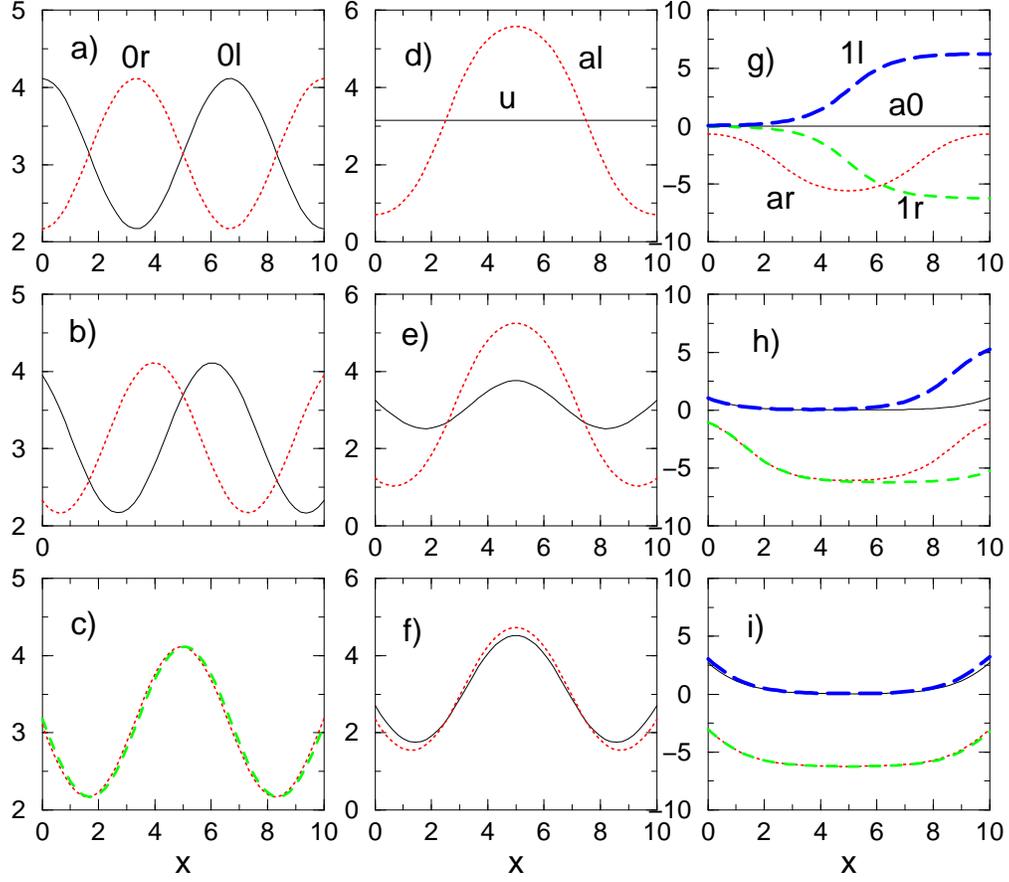,width=5.2in}
             }
  \caption{Plot of the phase
       distribution $\Phi (x)$ for all the solutions at $H=0$ and three 
values
        of the current (different for each line).
        ($a,~d,~g$) $I=0$,
       ($b,~e,~h$) $I=I_{max}/2$, ($c,~f,~i$) $I=I_{max}$ (where  
$I_{max}$
       is different for each case).  The three columns correspond to:
       ($a,~b,~c$) the two solutions $(0l,~0r)$ with $m^*=0.213839$.
       ($d,~e,~$f) the two solutions ($u$, $al$) from the left lobe of 
Fig.
        3b and
        ($g,~h,~i$) the two solutions ($1r,~1l$) for $m^*=0.999272$
        and the two solutions from the left lobe of Fig 5a  (i.e. 
$a0,~ar$)}
  \label{fig:6}
\end{figure}

We will present in more detail the calculations for $w=10$ since this
is a common length in experimental design of long junctions and one
can clearly see the multiplicity of solutions. In Fig.~\ref{fig:6} we present all
the solutions (for constant $x_0$ and constant $m$) for $H=0$ at three
different currents in order to follow their evolution. The branch with
a half period solution for $w=2K(m)$ (see $1r$ and $1l$ in Fig.~\ref{fig:6}g-i)
has a maximum current $I=4$ which is the same as the maximum current
in Fig.~\ref{fig:3}b. In fact at the maximum current (Fig.~\ref{fig:6}i) we have the
coalescence of four different solutions.  In the third column of
Fig.~\ref{fig:6} (i.e.  $g,~h,~i$) we see the four solutions being different at
$I=0$ (see $g$) but converging to the same solution (modulo $2\pi$) as
$I$ approaches the $I\approx 4$ (see $i$) value which is the maximum
current for all four solutions.  The four solutions come in pairs: two
from the pair with  $w=2K(m)$ discussed earlier (i.e.  $1r$, $1l$) and
two from the right lobe of Fig.~\ref{fig:3}b, i.e. $ar$ and $a0$ in
Fig.~\ref{fig:6}g-i, discussed in the previous subsection.   The other pair of
solutions with 3 half periods when $w=6K(m)$, i.e $0l$, $0r$ in
Fig.~\ref{fig:6}a-c have a maximum current near $I=1.8$. For higher currents
(above the value at  $c$) it jumps branch and converges to the
solutions of the left lobe of Fig.~\ref{fig:3}b since the two pairs of solutions
are quite close as can be seen from the plots in Fig.~\ref{fig:6}c and
\ref{fig:6}$f$.
Notice that the currents are different for the two plots. We should
also point out (to be discussed in the next section) that by
slightly increasing the magnetic field the $w=6K(m)$ solutions show an
interesting bifurcating behavior with a jump in the maximum tunneling
current. All solutions as seen in Fig.~\ref{fig:5}b come in pairs with opposite
fluxon content. Thus on the line $N_f=0$ (zero flux) we also have two
pairs of solutions up to $I=2.4$ (point $A$ in Fig.~\ref{fig:5}b, where only the
point at the maximum current is shown). One pair is the curves $u$,
$al$ in Fig.~\ref{fig:6}d,~e,~f. A second  pair goes up to $I=4.0$ (point
$B$), i.e. the curves    $a0$, $ar$ in Fig.~\ref{fig:6}g-i. The solutions of
each pair are different as can be seen in Fig.~\ref{fig:6}a,~d,~g or
Fig.~\ref{fig:6}b,~e,~h and only coincide at the maximum current. Notice that $u$
and $a0$ are simply displaced by $\pi$, while $al$ and $ar$ have
opposite signs.  With increasing current, though, they evolve very
differently.

\section{ No current case ($I=0$)}

In the absence of external current Eqs. (\ref{bca},\ref{bcb}) reduce to the following
\beq
2\sqrt{m} cn \left(x_0+\frac{w}{2} |m\right) 
=  H,~~~\label{bcI=0a}
\eeq
\beq
2\sqrt{m} cn \left(x_0-\frac{w}{2} |m\right) 
=  H, ~~~\label{bcI=0b}
\eeq

which determine the parameters $m
$ and $x_0$ that characterize the periodicity and phase shift of the
static solutions.
This is seen from Eqs. (\ref{bcI=0a},\ref{bcI=0b}) and the periodicities of
the elliptic functions.
Notice that these are the only three possibilities leading to
three branches, if we consider solutions, where only $m$ varies and
$x_0$ is fixed to $x_0=0$ or $x_0=2K(m)$. Here we need $x_0=2K(m)$,
and not $x_0=-K(m)$ as in the zero field solution, due to the
symmetric boundary conditions  for the magnetic field. At the same
time we also have solutions where $m$ is fixed and $x_0$ is varied
continuously with the current.

\subsection{Fixed $x_0$ solutions}

We start by giving the three   branches.

{\bf Branch I:} An obvious choice in (\ref{bcI=0a},\ref{bcI=0b}) is $x_0=0$, so
that for $m \le 1$
\beq  H  =  2\sqrt{m}~ cn
(\frac{w}{2}|m) ,~~~ 0\le m <1,\,\,\,~~~\label{BIa}
\eeq
\beq
\cos \Phi_0(x)  =  2~{m}~ sn^2 (x|m)-1 ,~~~
0\le m <1,\,\,\,~~~\label{BIb}
\eeq
\beq
\sin \Phi_0(x)  = 
-2~\sqrt{m}~ sn (x|m) dn (x|m),~~~ 0\le m <1.\,\,\,~~~\label{BIc}
\eeq

Another possibility is $x_0=2K(m)$ for the
case that $K(m)~<~ \frac{w}{2}~<~3K(m)$, or more generally when $cn
(\frac{w}{2}|m)<0$, since we are limiting ourselves to $H>0$. This
means that every time $w$ increases by $2\pi$ we introduce two extra
solutions. To put it in other words the function  $H(m)$ in (\ref{BIa})
 is highly oscillating for large $w$.

{\bf Branch II:} For $x_0=0$  and $m>1$ we use the notation
$\bar{m}=1/m$ and the transformation rules of elliptic functions \cite{AS} to
obtain
\beq H  =  \frac{2}{\sqrt{\bar{m}}}~ dn
\left(\frac{w}{2}\frac{1}{\sqrt{\bar{m}}}|\bar{m}\right) ,~~~ 0\le
\bar{m} <1,\,\,\, ~~~\label{BIIa}
\eeq
\beq
 \cos \Phi_0(x)  =  2~ sn^2
\left(x\frac{1}{\sqrt{\bar{m}}} |\bar{m}\right)-1 ,~~~ 0\le \bar{m}
<1,\,\,\, ~~~\label{BIIb}
\eeq
\beq
 \sin \Phi_0(x)  =  -2~ sn
\left(x\frac{1}{\sqrt{\bar{m}}} |\bar{m}\right) cn
\left(x\frac{1}{\sqrt{\bar{m}}} |\bar{m}\right),~~~ 0\le \bar{m}
<1.\,\,\,~~~\label{BIIc}
\eeq

{\bf Branch III:} Taking into account the period and
symmetry of the $nd$ elliptic function  we can also put in Eqs.
(\ref{bcI=0a},\ref{bcI=0b}) $x_0=\frac{1}{\sqrt{m}} K(\frac{1}{m})$ with $m>1$  (it
cannot be satisfied for $m<1$) to  obtain
\beq H  =  2
\sqrt{\frac{1-\overline{m}}{\overline{m}}}~ nd
\left(\frac{w}{2}\frac{1}{\sqrt{\bar{m}}}|\bar{m}\right) ,~~~ \bar{m}
<1,\,\,\,~~~\label{BIIIa}
\eeq
\beq
 \cos \Phi_0(x)  =  2~ cd^2
\left(x\frac{1}{\sqrt{\bar{m}}} |\bar{m}\right)-1 ,~~~
\,\,\, ~~~\label{BIIIb}
\eeq
\beq
 \sin \Phi_0(x)  =  2~ \sqrt{1-\bar{m}} \,\,cd
\left(x\frac{1}{\sqrt{\bar{m}}} |\bar{m}\right) sd
\left(x\frac{1}{\sqrt{\bar{m}}} |\bar{m}\right).\,\,\,~~~\label{BIIIc}
\eeq
The branches II and
III were obtained from Eqs. (\ref{bcI=0a},\ref{bcI=0b}) by assuming that the modulus $m >1 $
and putting $\bar{m}=\frac{1}{m}$. The expressions (\ref{BIa}) and 
(\ref{BIIa}) can
both be written in the form 
\beq 
H=2\sqrt{m}~ cn (\frac{w}{2}|m) ,~~~
0\le m <\infty .\,\,\,~~~\label{eq20} 
\eeq 
In this case branch  II is
described by Eq.  (\ref{eq20}) at $m \ge 1$  which reduces to 
(\ref{BIIb}) by
using the transformation formulas (see Ref. \cite{AS} ) when the
modulus is greater than unity.   Again for large $w$,  $H(m)$ is
strongly oscillating, even though it remains always positive.

In Fig.~\ref{fig:7} we plot the fluxon content $N_f$, at $I=0$ for three
lengths, that show different patterns and evolution
and the introduction of multiple solutions with length. In the first
row we plot the magnetic field ($H$ vs $m$) and we see that for the same $H$ we
have different flux.  This means that they correspond to different
solutions with different screening currents.
Comparing the second and the third rows we see
that the modulus $m$ (at zero current) is a much better parameter than
the magnetic field $H$ to characterize the solutions, since the flux in
this case is unique except for a symmetry multiplicity. Also notice
that the curves should be symmetric about the horizontal at the zero
level (two $x_0$ values), but the plot is not completed to keep the
vertical scale shorter and avoid optical complexity to the eye.  Only
 for branch I we show both curves (for $x_0=0$ and $x_0= 2K$). On the
other hand the magnetic field plots exhibit some interesting changes
of slope which for very long junctions alternatively correspond to
stable and unstable regions of solutions. The change of slope is
especially apparent for $w=\frac{5\pi}{2}$ but the alternation of
stable and unstable regions also exists for small lengths as will be
discussed later.

Notice
that at the points of slope change the fluxon content is an integer
for both branches II (light dashed) and III (dark dashed). Actually
for stronger $H$ the curves in Fig.~\ref{fig:7}i will look just like in
~\ref{fig:7}c.
Notice also that the symmetry in figures (b),(e),(h) will correspond
to an antisymmetric form in the plot of flux ($N_f$) with $H$. The
 oscillatory form of flux vs $H$ is understood by looking at the
 relation of $H$ and $m$ as plotted in (a),(d) and (g) for the three
 lengths. Notice the evolution with the creation of lobes for small
 $m$, whose number will increase with $w$ as discussed.  In the
 $w=\frac{5\pi}{2}$ case we also have extra solutions with $m$ fixed
 which are not shown in the plot, but will be  discussed in what
 follows.  Also for higher $w$ the $N_f$ plot becomes more complex and
as a particular case we discuss the $w=10$ length in the next
subsection.

\subsection{Magnetic flux  for $w=10$ }

 In Fig.~\ref{fig:8} we plot the fluxon content $N_f$, for a junction length
$w=10$, at zero current as a function of, the magnetic field $H$ in (a)
and equivalently the modulus $m$ in (b). We see that the plot in
(b) is essentially single valued, while the two curves correspond to
the choices $x_0=0$ and $x_0=2K(m)$ (of branch I), which give solutions
with opposite flux. The corresponding plot with $H$ is quite deformed
(due to the periodic relation between $H$ and the modulus $m$). In the
plots the lines are the results of the analytic solutions and the
 symbols the numerical simulation results.
In the second case we have to try different runs to
 complete the curve. 
As we see the plot of $N_f$
vs $m$ can be continuously traced by varying the modulus $m$ in the
analytic solutions. Then one can obtain the magnetic field $H$ from $m$
using analytic expressions (shown in Fig.~\ref{fig:8}c) and also trace the
curve in (a). When, however you do numerical simulations, using the
magnetic field as a varying parameter, you can trace only the part of
the curve with the same slope. When you reach an extremum in the
magnetic field (see points $g,\, l,\, i,\,k,\,r,\,q, \cdots$ in
Fig.~\ref{fig:8}a),
where the slope becomes infinite, the iteration procedure for
the branch continuation with increase of $H$ does not converge. Then
you must decrease the value for $H$ to trace the negative slope curve.
Thus one needs five tracings back and forth in $H$ to obtain all the
solutions of branch $I$, i.e. the curves $0l$, $0r$, $u$, $1l$, $1r$
in Fig.~\ref{fig:8}a.

In Fig.~\ref{fig:8} we show all three branches : branch I as given by the
curves $o-i-g-e$  for $x_0=0 $ (solid line) and $o-k-l-m)$ for
$x_0=2K(m) $ (dotted line); branch II is given by the line
$e-c-a-\cdots $ (long dashed line); and branch III is given by the line
$ o-r-q-p \cdots $ (dashed line).  Near $H$=0 for I=0 we have several
solutions four of which have the same fluxon content $N_f =0$, i.e.
$al$, $ar$, $a0$, $u$, and four with different $N_f $ at points $
e,m,x$ and $y$, i.e. $1l$, $1r$, $0l$, $0r$. Notice that the point $e$
is slightly to the right of the $N_f $ axis. This is because, for the
particular value of $w=10$ near $m=1$,  the $cn(\frac{w}{2},1)$
elliptic function behaves like ${\rm sech}(\frac{w}{2})$ so that for large
$w$ it is small and positive.

When we increase the current $I$ and fix
$H=0$, the points $x,y$ of Fig.~\ref{fig:8}, will give us the curves $0r$ and
$0l$ in Fig.~\ref{fig:5}b for the flux. The corresponding points in the
neighborhood of $e$ and $m$ will give us the curves $1l$ and $1r$
in Fig.~\ref{fig:8}. At the point $o$ there are two solutions with constant $m$,
i.e. $al$, $ar$ in Fig.~\ref{fig:8}a (which for  $H=0$ are part of the
curve $N_f =0$). With increasing current one of them (i.e. $al$)
reaches the maximum current at point $A$ in Fig.~\ref{fig:5}b with $I=2.5$. The
other (i.e. $ar$) goes to B in Fig.~\ref{fig:5}b with $I\approx 4.0$. These
solutions have $m=0.88299$ where $cn (w|m)$ for the given $w$
vanishes.  They have opposite flux because they correspond to the two
possible values of $x_0=-K(m),K(m)$. They are equivalent to the
solutions in the middle point  with $I=0$ in Fig.~\ref{fig:3}b.  In the point
$o$ there are two more solutions from which one belongs to the stable
 branch $o-r$ (i.e. $a0$) and the other to the unstable branch $i-o-k$
 (i.e. $u$) in Fig.~\ref{fig:8}a.

\subsection{Fixed $m$ solutions}
 The first branch (I) has also solutions with fixed $m=m^*=0.88299$ if
$w>2\pi$.  The value of $m^*$ is obtained from the condition
$\frac{w}{2}=2K(m^*)$, so that we have an integer number of periods
$(4K(m))$ for the elliptic function in the junction length $w$. This
automatically satisfies the symmetry requirement for the boundary
conditions. The magnetic field is determined from the position phase
parameter $x_o$ (with $H$ a periodic function of $x_o$ with period
$4K(m^*)=w$). It is given by $$H=2\sqrt{m^*}cn(x_o+2K(m^*)|m^*)$$ and
is presented in Fig.~\ref{fig:9} for $w=10$. The maximum value of $H$ for this
branch is $H=1.8$ at $x_o=\pm 2K(m^*)$. The two signs correspond to
the $al$ and $ar$ curves. Notice that these solutions at zero current
have zero fluxon content ($N_f=0$) over the whole extent of the
magnetic field for which they exist. These solutions  also exist for
$w=\frac{5}{2}\pi$, but are not shown in Fig.~\ref{fig:7}h,~i.  For larger $w$
we expect more pairs of solutions, i.e. a pair for each increase of
$w$ by $2\pi$.  Thus for $4\pi<w<6\pi$ there are two pairs of
solutions.
\subsection{Maximum tunneling current}
In Fig.~\ref{fig:10} we plot the maximum tunneling current as a function of the
magnetic field for the three different lengths. We see that for
$w=\frac{2\pi}{3}$ there are two curves (like $a_0$ and $u_0$ in
(a)) for the maximum current, one of which is stable and the other
unstable. There are,  however, abrupt variations of the maximum
current. Thus at the end of the stable (0-1) branch (line $a-b$) there
is an unstable (1-2) branch (line $c$-$d$), which however has a
discontinuity at about $H=3.3$ in $I_{max}$ (see vertical arrow). The
 same happens at the left of the  stable (1,2) branch $d-n-c-m$ where
its continuation  has a discontinuity of 0.4 in $I_{max}$, at about
$H=2.6$. We see that at the $I_{max}$ there are two curves
superimposed with the same $I_{max}$ but they start from different
solutions at $I=0$. Thus one has to be very careful when tracing
numerically the $I_{max}$ vs $H$ curve.  For intermediate current
 values $(I<I_{max}$ at a fixed $H$) you must check to the right of
$c$ which solution branch you follow. Thus if we trace for the
$I_{max}$ the stable branch at $I=0$ from $H=0$ (i.e.  $\Phi_0(x)=0$)
to the right we follow the curve $a-p-b-c-n-d ~\cdots $, while if we
 start from the unstable branch at $H=0$ (i.e.  $\Phi_0(x)=\pi$) we
follow the curve $a-p-m-c-n-d \cdots $.

\begin{figure}
  \centerline{
     \psfig{figure=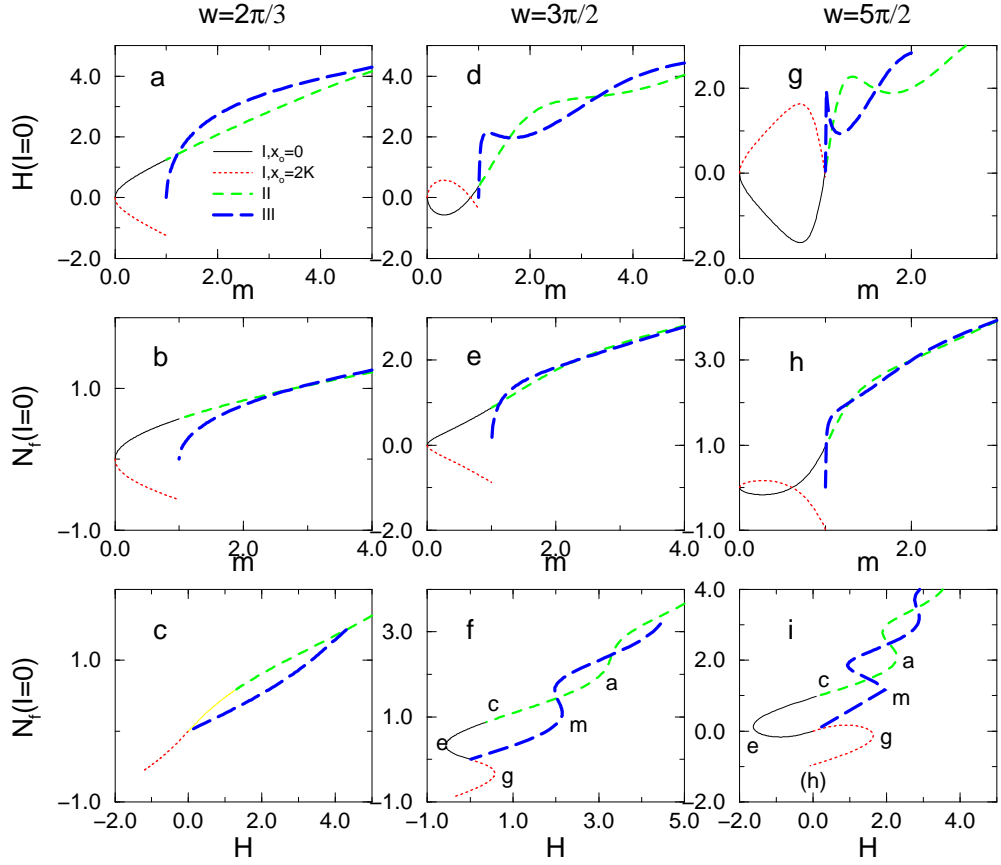,width=5.2in}
             }
  \caption{Plots of $H$ vs $m$ (top figures) and of the fluxon
content $N_f$ for $I=0$  as a function of $m$ (middle figures) and
$H$ (bottom figures), for a junction of length $w=\frac{2\pi}{3}$,
 $\frac{3\pi}{2}$, $\frac{5\pi}{2}$. }
  \label{fig:7}
\end{figure}

\begin{figure}
  \centerline{
     \psfig{figure=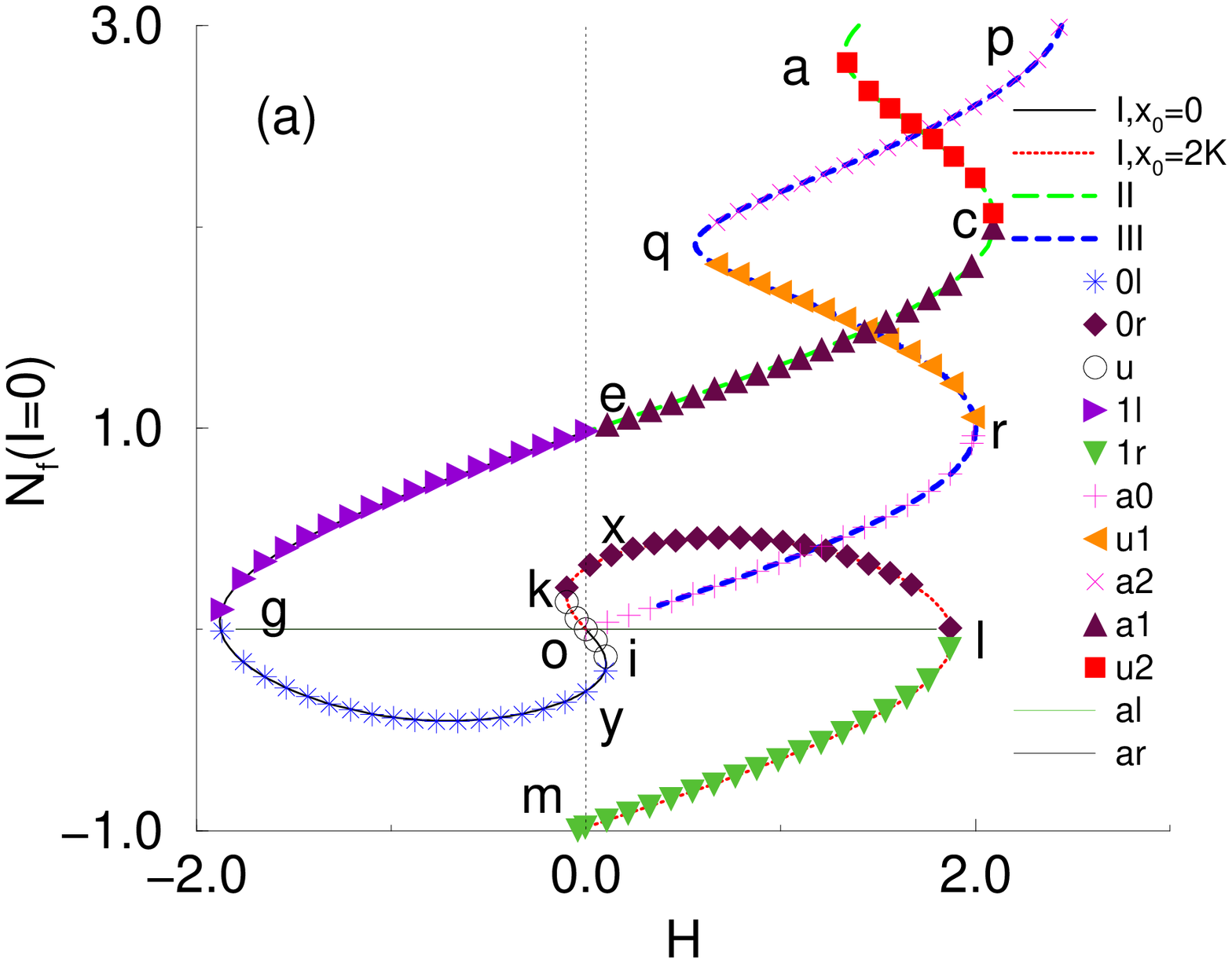,height=2.3in}
             }
  \centerline{
     \psfig{figure=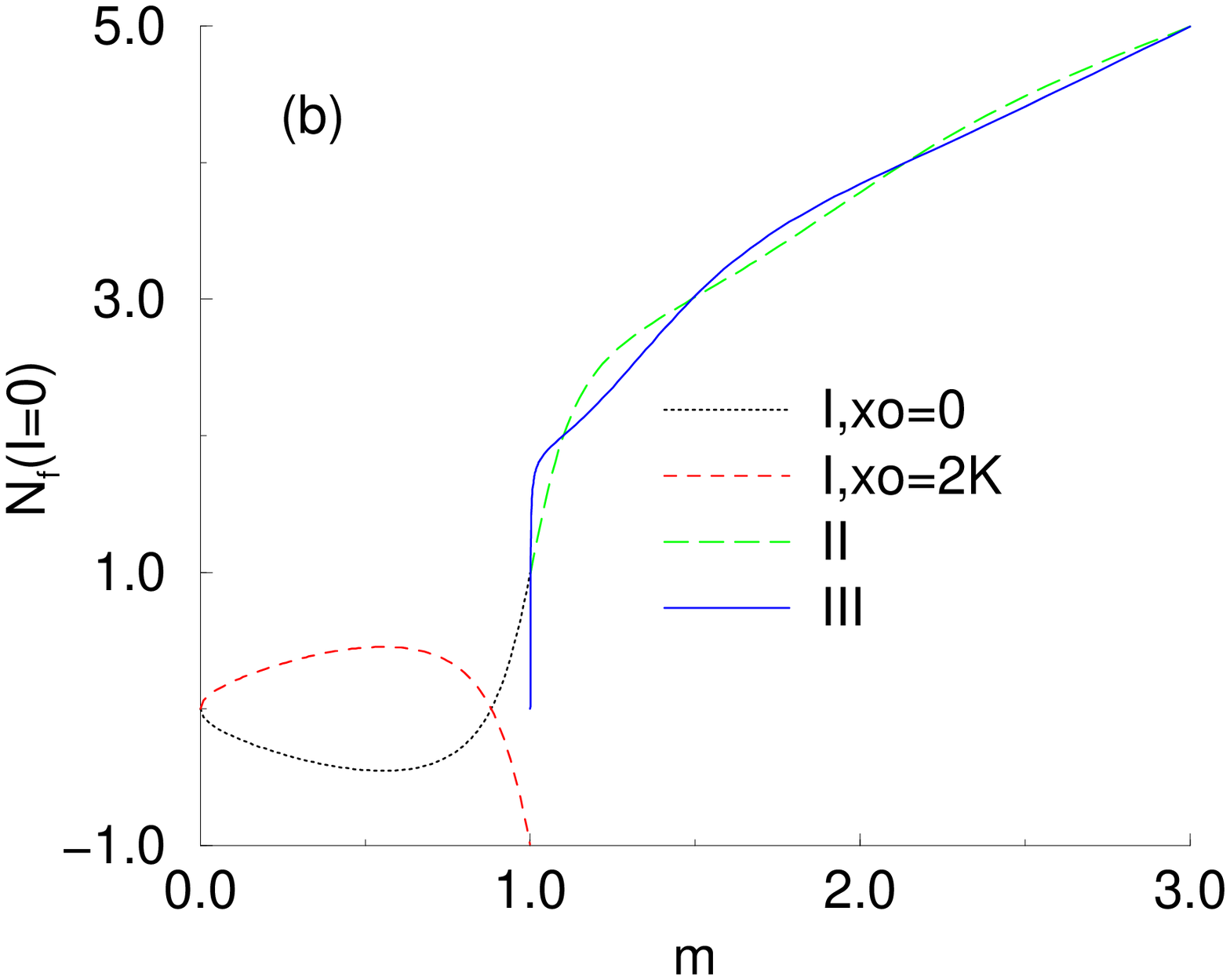,height=2.3in}
             }
  \centerline{
     \psfig{figure=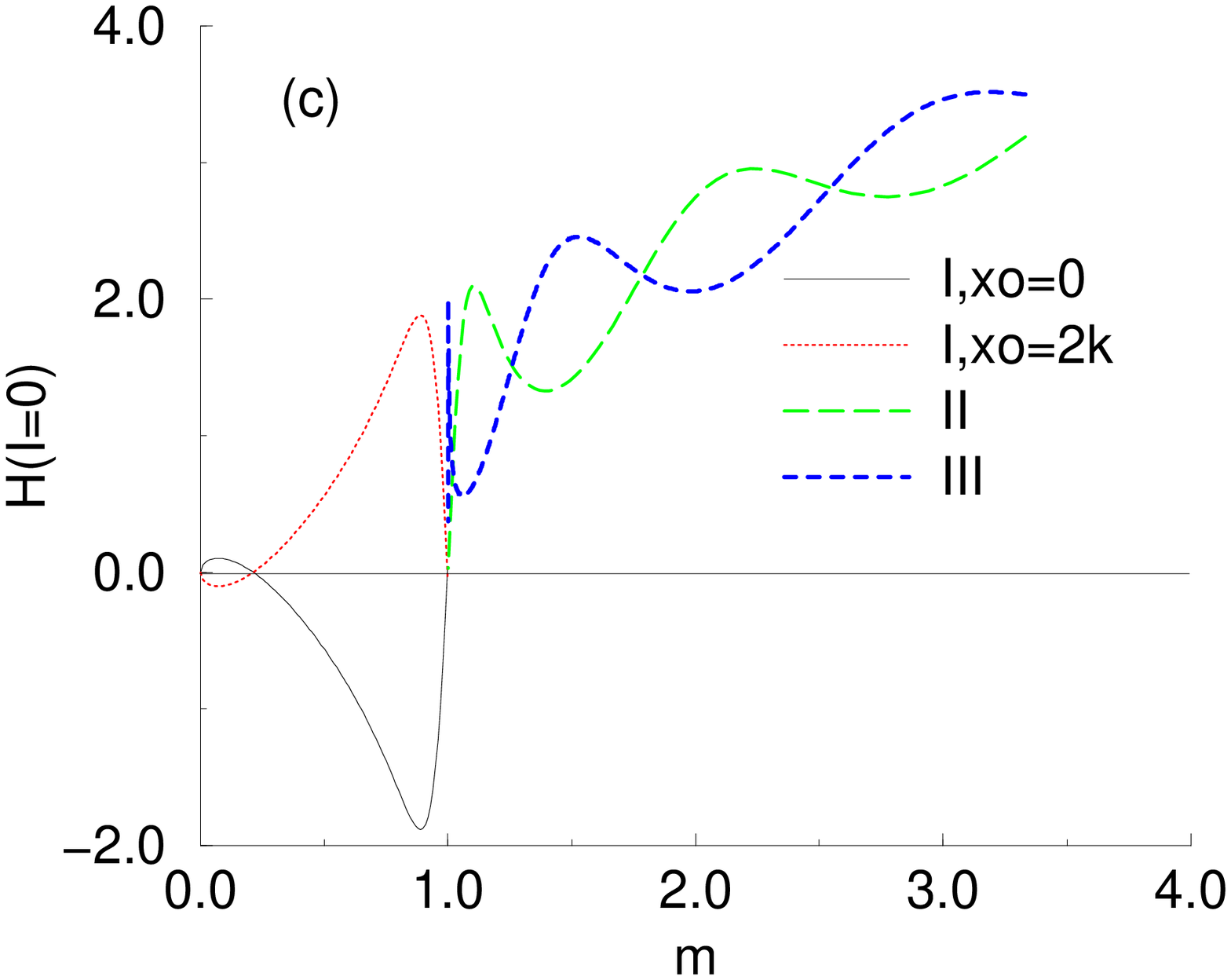,height=2.3in}
             }
  \caption{Plot of
     the fluxon content $N_f$ for $I=0$ and $w=10$ as a function of (a)
     $H$ and (b) $m$.  In (c) we give $H(m)$.}
  \label{fig:8}
\end{figure}

\begin{figure}
  \centerline{
     \psfig{figure=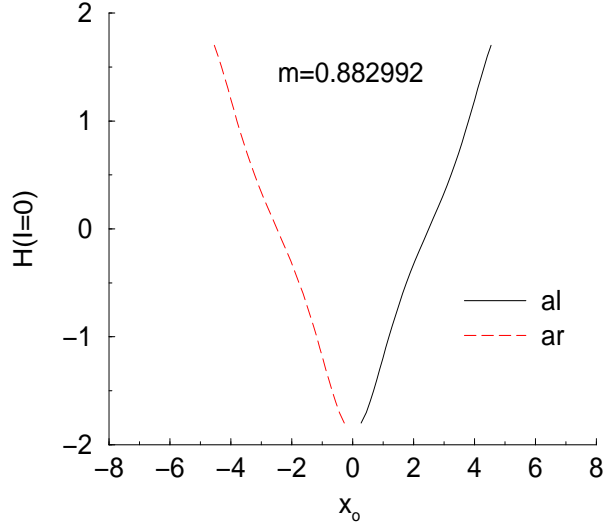,height=2.8in,width=3.1in}
             }
  \caption{Magnetic field $H(x_0)$ at $I=0$ for the constant 
$m=0.882992$
     solution.}
  \label{fig:9}
\end{figure}

\begin{figure}
  \centerline{
     \psfig{figure=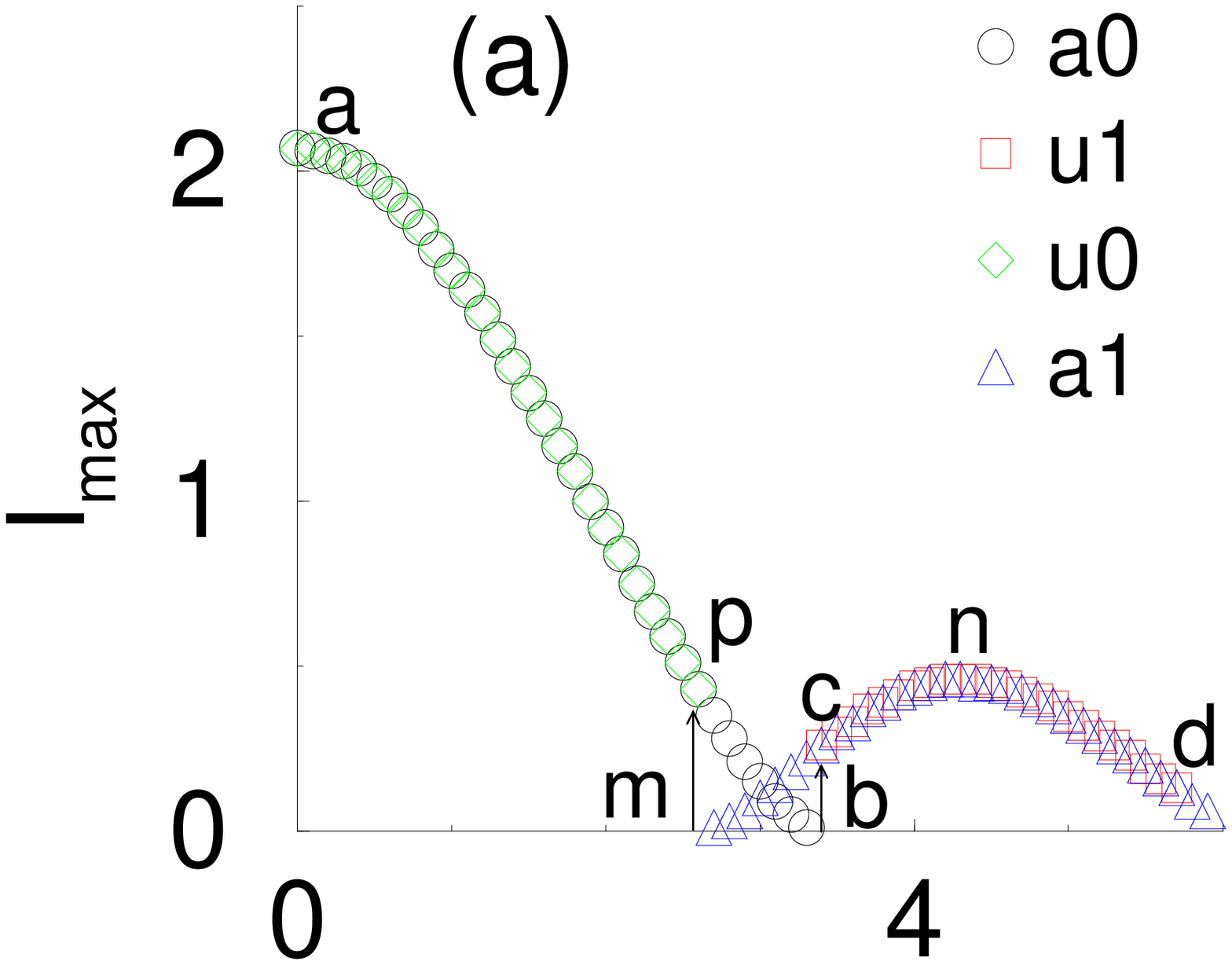,width=1.7in}
     \psfig{figure=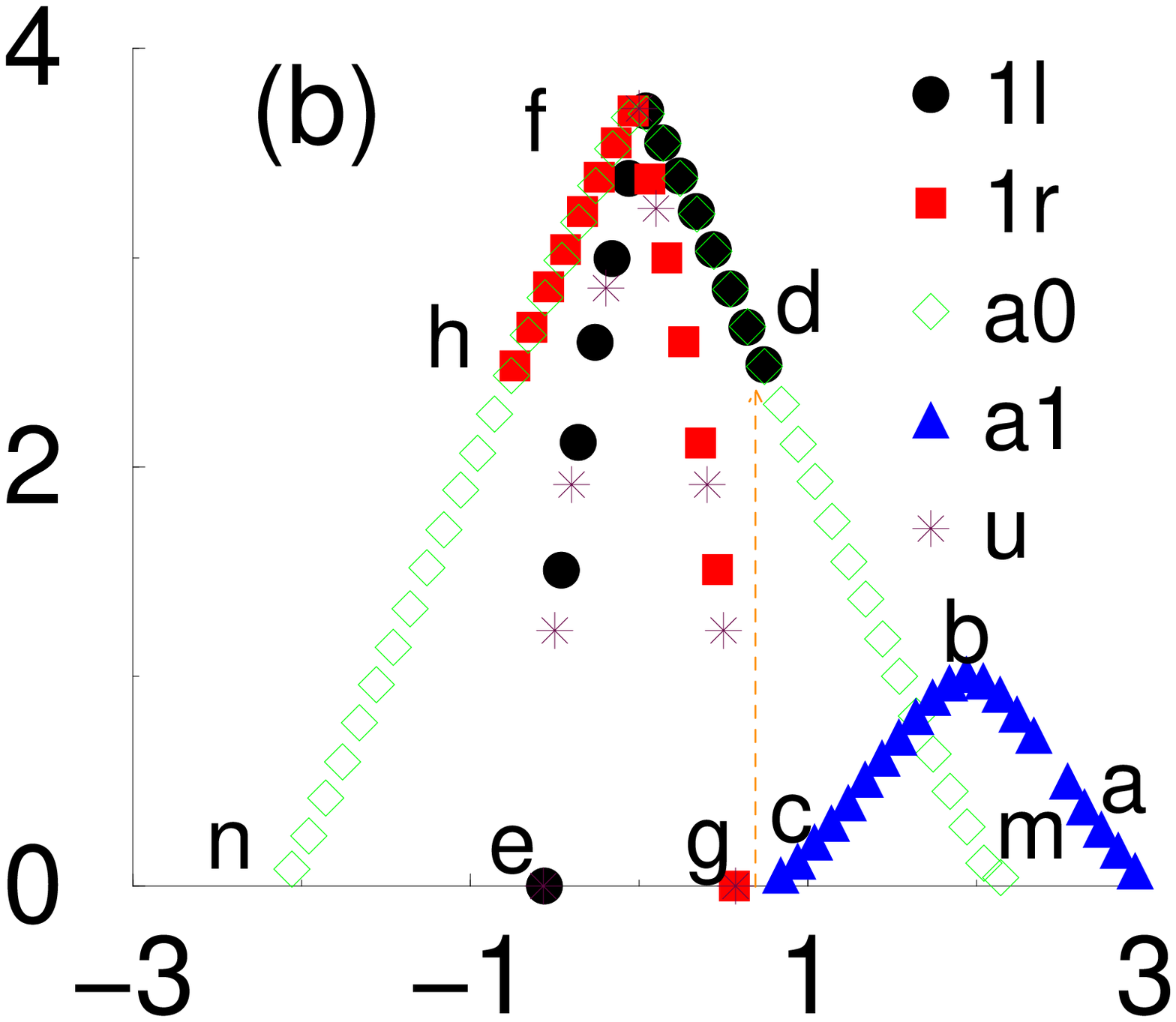,width=1.7in}
     \psfig{figure=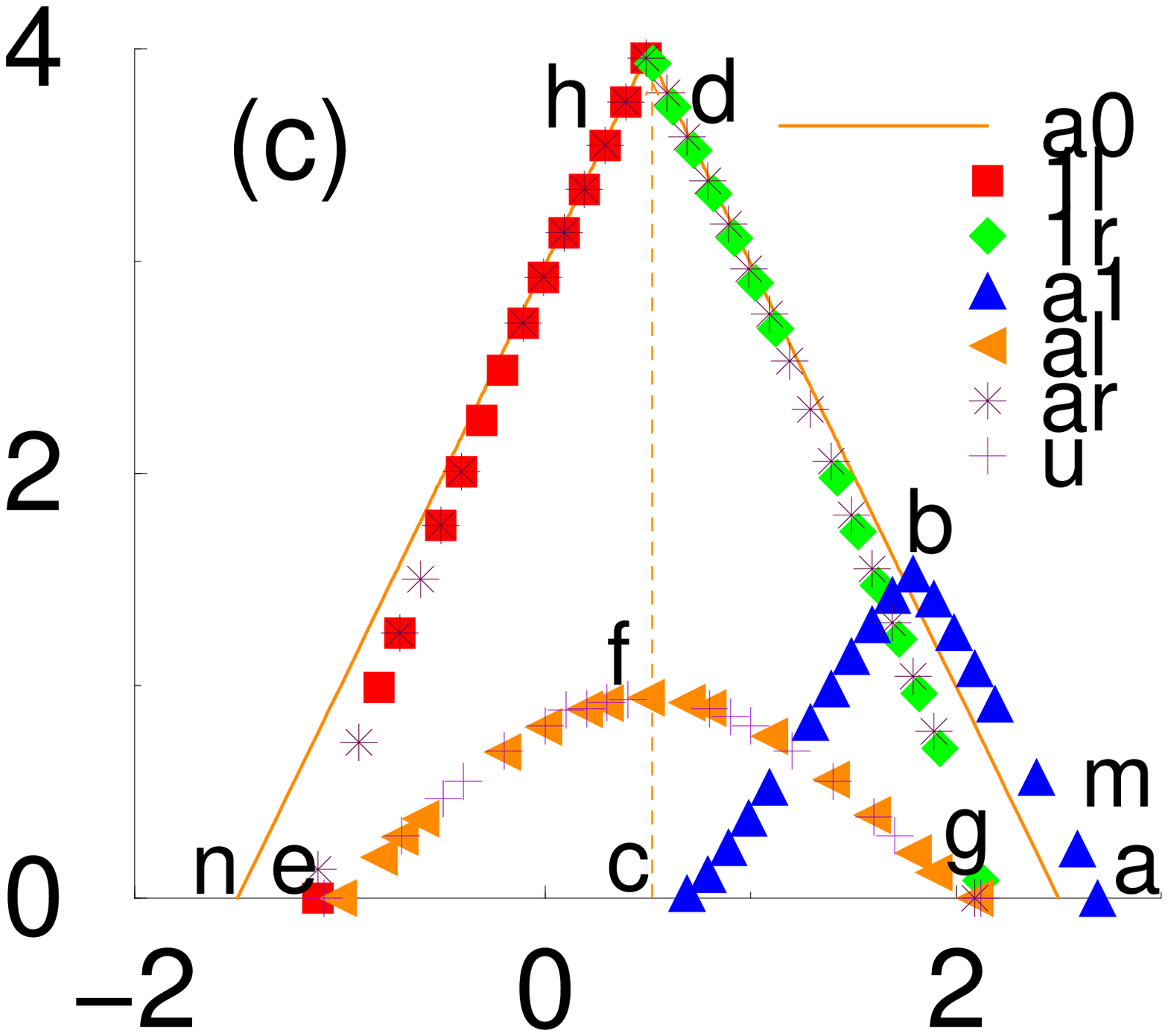,width=1.7in}
             }
  \centerline{
     \psfig{figure=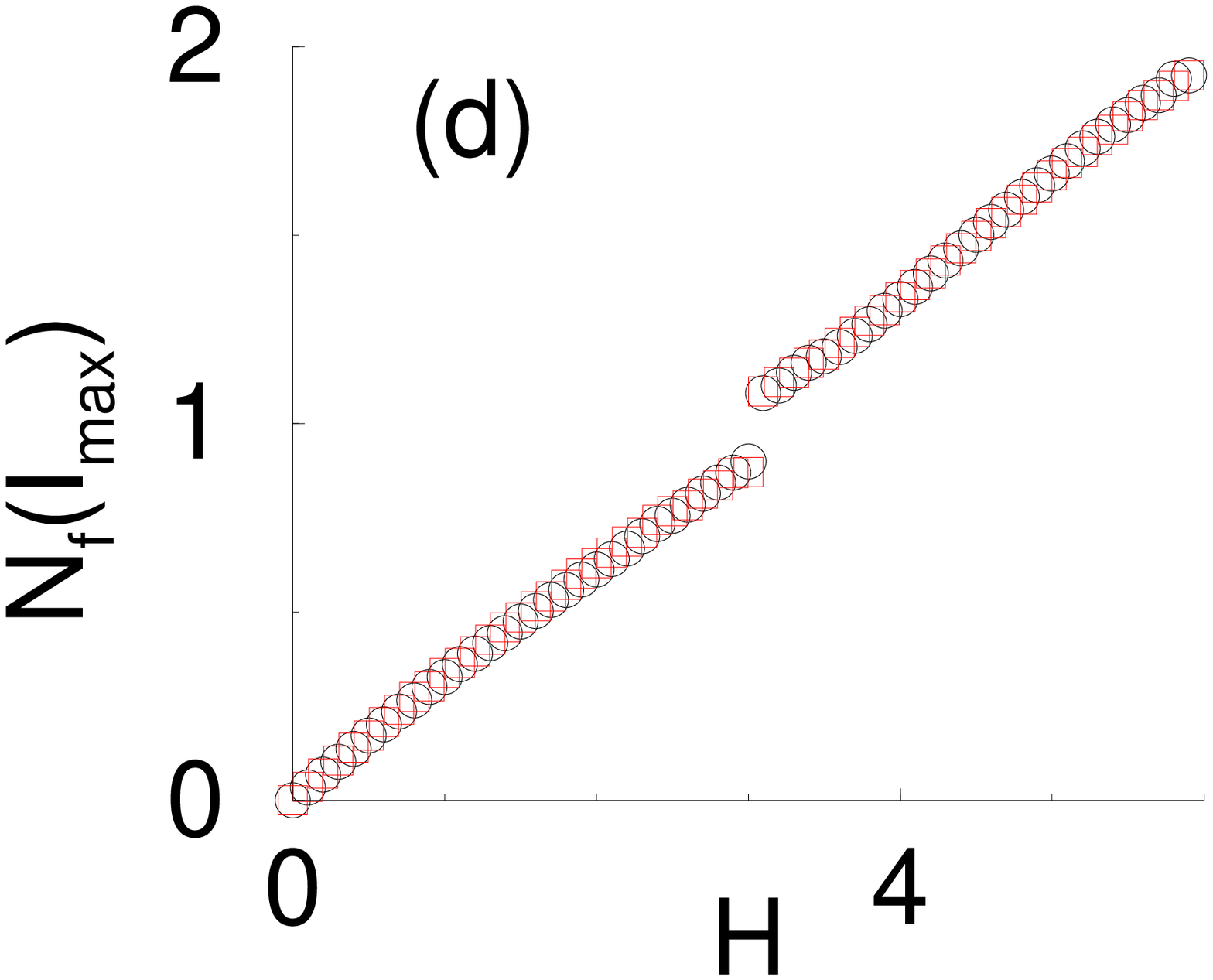,width=1.7in}
     \psfig{figure=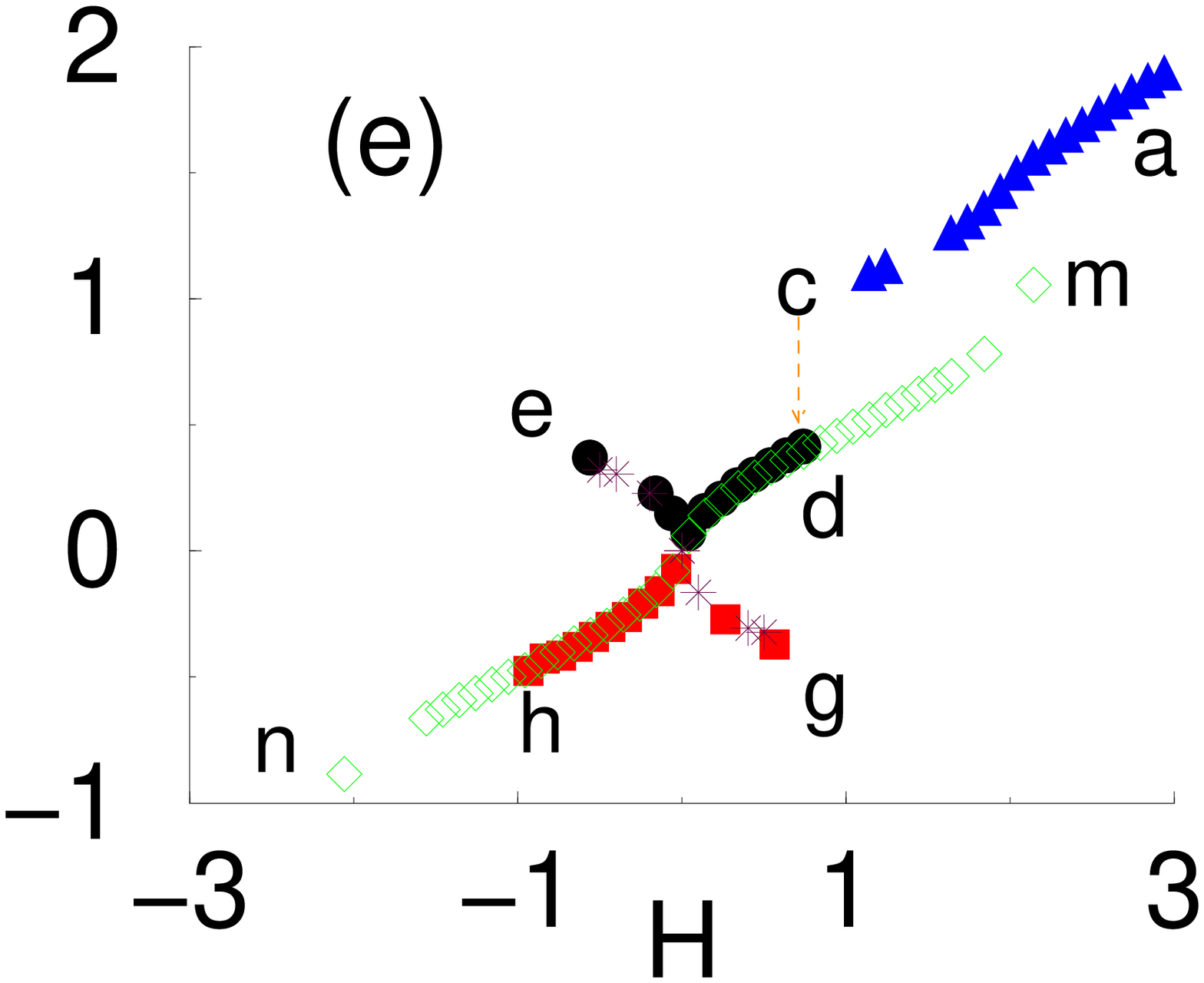,width=1.7in}
     \psfig{figure=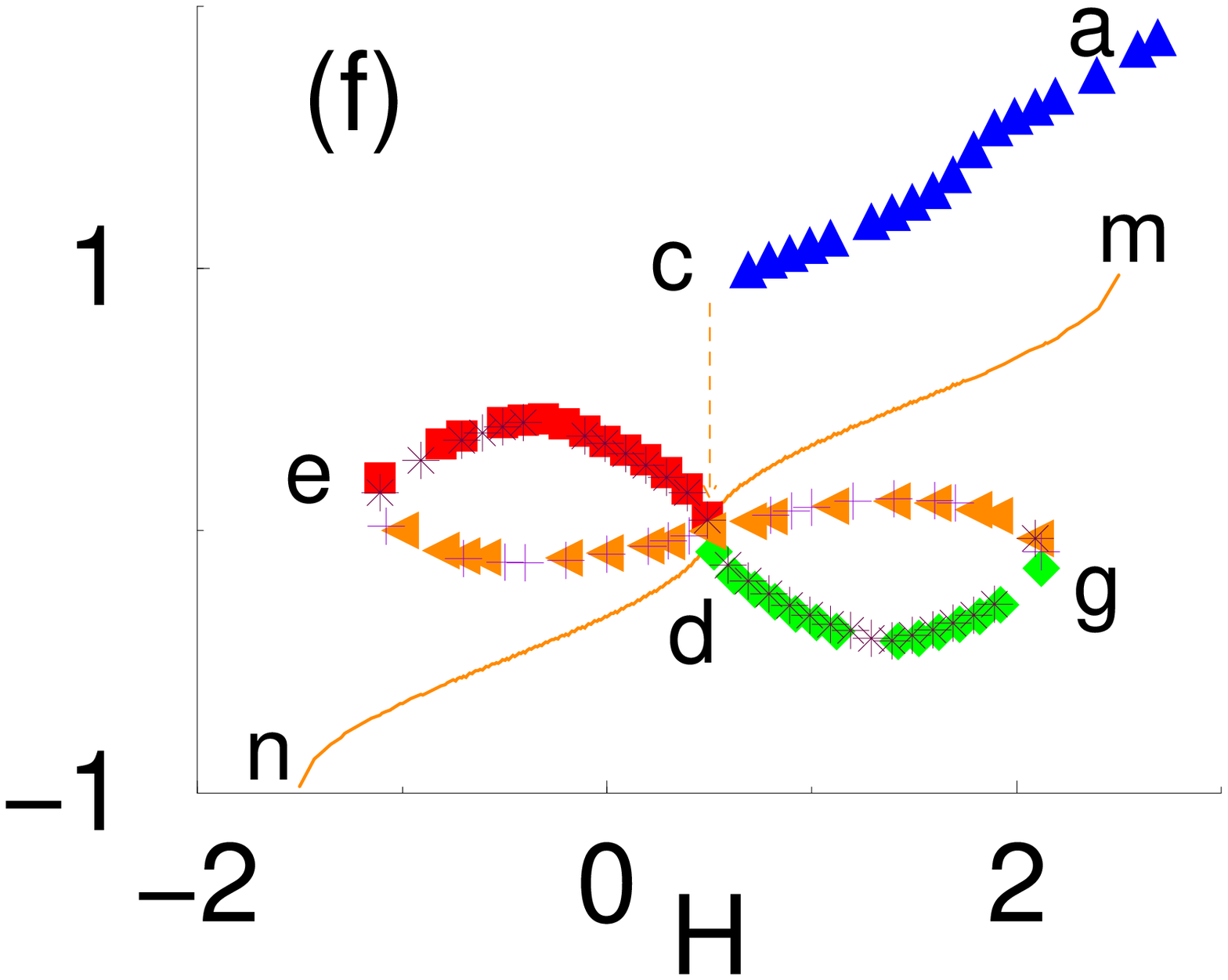,width=1.7in}
             }
  \caption{Maximum tunneling current as a function of $H$ for three 
different
     lengths (a) $w=\frac{2\pi}{3}$, (b) $w=\frac{3\pi}{2}$, (c)
     $w=\frac{5\pi}{2}$ and $N_f$ vs $H$ at the maximum current 
(d),(e),(f)
     correspondingly for the three lengths.}
  \label{fig:10}
\end{figure}

The plot for
$w=\frac{3\pi}{2}$ (Fig.~\ref{fig:10}b) is a bit more complicated. This to some
extend is caused by the loops in Fig.~\ref{fig:7}d. Thus when we trace from
$H=3.0$ to the left, we follow the curve $a-b-c$ with a jump to $d$
then $d-f-e-f-g-f-h$, followed by a jump to a point symmetric to $c$
and from then on following a symmetric path which is not shown in the figure.
The corresponding path for $w=\frac{5\pi}{2}$ (Fig.~\ref{fig:10}c) is $a-b-c$ with a
jump to $d$ then $e-f-g$, to $h$ and from there on a symmetric
curve. Notice that in this case, the extra branch $e-f-g$ has a lower
peak current from the previous cases.

In  the above $I_{max}$ diagrams the existence of jumps as we scan the
magnetic field  and  increasing the $I$ value (at fixed $H$) implies a
dependence of the final solution at $I_{max}$ on the initial condition
and the path of approaching it. This becomes clear if we connect it
with the morphology of the $I$ and $H$ contours in Fig.~\ref{fig:2} for the case
$w=3\pi/2$.

As we see there are four paths to reach the point $f$ (notice
corresponding points in both Figs.~\ref{fig:2} and ~\ref{fig:10}b) 
if we fix $H=0$ while
increasing the current. These paths are along the curves $x_0=3K(m)$
(which is equivalent to $-K(m)$) and the vertical line $m=m^*=0.84$.
From these only the one from the left of $f$ along  $x_0=3K(m)$ (up
to $m=1$) is stable. This corresponds to the solution $a_0$. The other
three paths will give solutions $u$, $1l$, and $1r$. The last two are
along the vertical line and $u$ along $x_0=3K(m)$ from $O$ to $f$.
Notice that the whole vertical axis (i.e. $m=0$) corresponds to the
single point ($H=0$, $I=0$) in the $I$-$H$ diagram.

 From the contours of $H$ around the points $e$ and $g$  it is
clear that at these points we have extrema of $H$ which also fall
on the curves $x_0=4K(m)$ and $x_0=2K(m)$ where $I=0$. Let us take
another look in Fig.~\ref{fig:10}b. If we start from the point $a$
(with $N_f=2$) by decreasing $H$ at $I=0$ we reach the point $c$
(with $N_f=1$) along the line $x_0=0$ (i.e. branch II) in
Fig.~\ref{fig:2}. The continuation of branch II through $m<1$ is
branch I which goes up to point $e$ in Fig.~\ref{fig:10}b. In the
rest of the curve, i.e. when $m$ goes from $e$ to $O$, the
magnetic field is reversed from -0.6 (at $e$) to zero. The range
in $H$ from $c$ to $m$ corresponds to two different paths in the
$x_0-m$ diagram. From the topology it is clear that they end up in
different maximum current as $H$ is kept constant. The path $mc$
(above $m$) has its maximum current to the left of the line, while
the path $cm$ (i.e. below $m$) has its maximum current to its
right. Notice that  in Fig.~\ref{fig:10}b, in going from $a$ to
$c$ (along branch II) you cross the $H$ value at $m$. The
corresponding point in the $x_0$-$m$ diagram is a different one
($m'$) on the $m$-axis between $a$ and $c$. Finally the point $b$
can be reached from several paths. A similar
 analysis can be given in all cases, but we chose a single length as a
point of illustration.

\begin{figure}
  \centerline{
     \psfig{figure=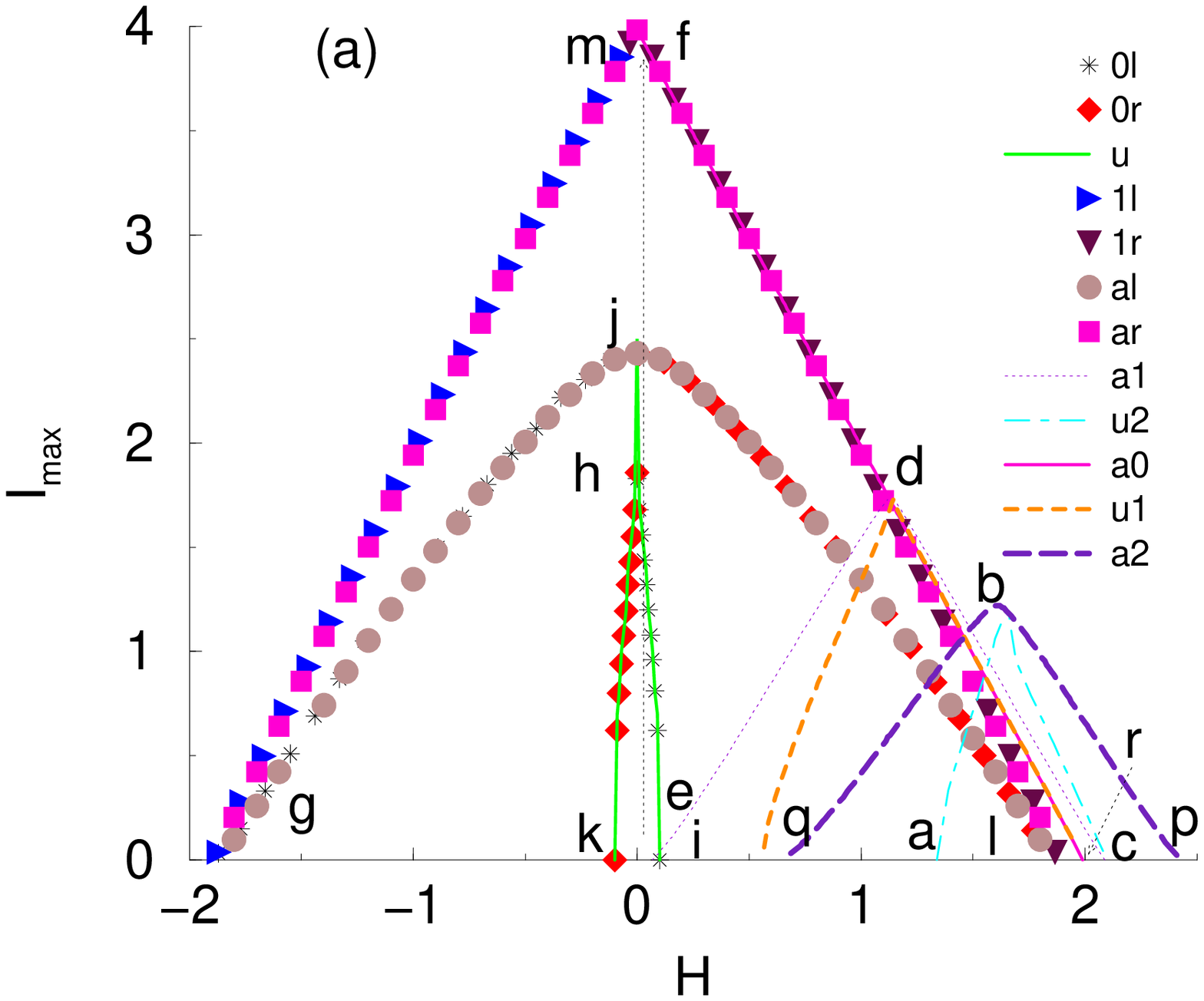,height=2.85in,width=3.0in}
             }
  \centerline{
     \psfig{figure=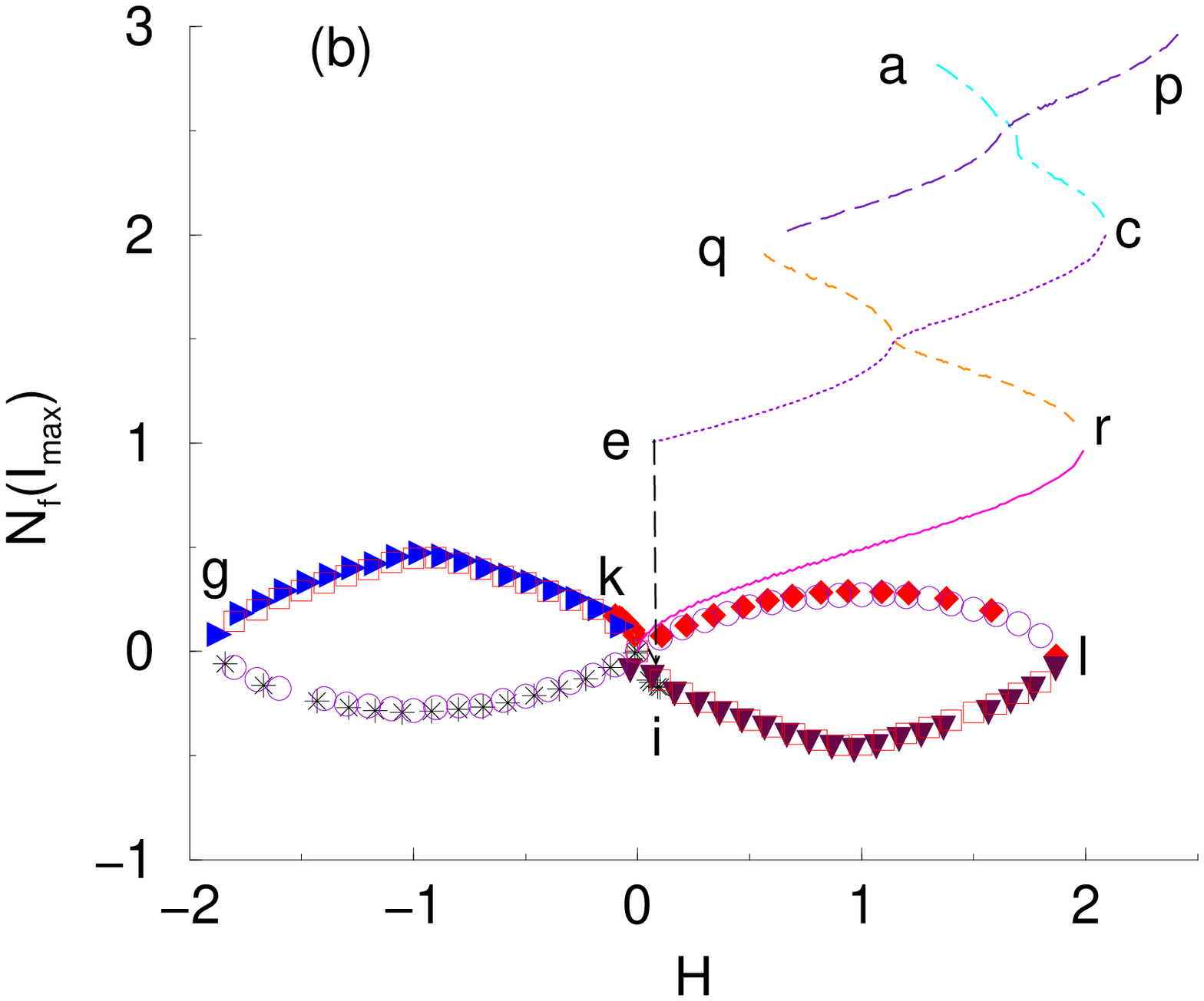,height=2.85in,width=3.0in}
             }
  \caption{The same as Fig. 10
     for $w=10$. (a) $I_{max} (H)$, (b) $N_f (H)$ at $I=I_{max}$.}
  \label{fig:11}
\end{figure}

Below each case in Fig.~\ref{fig:10} we plot the
corresponding flux vs $H$ at the maximum current. These curves
resemble the ones at zero current in Fig.~\ref{fig:7}(c,~f,~i). One remark
is that while at $I=0$ the $N_f$ vs $H$ is a continuous curve, at
 $I_{max}$ the flux shows discontinuities the same way that the
maximum current was discontinuous. Comparing Figs~\ref{fig:7}f
and ~\ref{fig:7}e we
see that the branches $om$ ($o$ being the origin) and $ca$ are only
slightly modified. The branch though $e-c$ (in Fig.~\ref{fig:7}f) is folded
onto $e-o-d$ (in Fig~\ref{fig:10}e). The same happens for $w=\frac{5\pi}{2}$
between Figs.~\ref{fig:7}i and 10f. In the plot for the flux we have not
 plotted all the branches.  For a more complete plot of the branch II
 and III solutions see the case $w=10$ in Fig~\ref{fig:11} since they have the
 same approximate structure.

 In Fig.~\ref{fig:11}a we show the maximum tunneling
current and the flux (at $I_{max}$) as a function of the magnetic
field, for $w=10$.  At this length we are already at the limit of
long length branches.  In this case when going from right to left we
trace the $I_{max}$ through the points $a-b-c-d-e$, jump to
$f-g-h-i-j-k-l-m$ with a symmetric continuation. The flux (Fig.~\ref{fig:11}b) also
shows a similar folding as for the case $w=\frac{5\pi}{2}$. The
letters correspond to the ones in $I_{max}$ vs $H$ plot.

\section{Analytical stability analysis}

Next we  obtain some analytic estimates
for the stability regions for the zero current solution.
 The observations obtained by these estimates will  verify
and extend our results by  solving numerically the stability
eigenvalue problem in (\ref{eq10}).  Substituting Eq. (\ref{eq04}) into Eq.
(\ref{eq10}) we obtain the Lam\'{e} eqn.  \beq -X'' + \left[2m ~sn^2
(x+x_0|m)-1\right] X =\lambda ~X.  \label{eq21} \eeq  Numerical
solution of (\ref{eq21}) with the boundary conditions in (\ref{eq11})
gives us all the eigenvalues and we can check the stability of the
static solution. One can gain some insight into a necessary bound for
stability from the following analytic considerations, which  we
also compare with the numerical results.

Let us remark that for three values of $\lambda =m-1, 0, m$  we can
give an explicit analytic form for the corresponding solutions of
(\ref{eq21}), which are 
\beq X_0= dn (x+x_0|m),\,\,\,
\lambda_0=m-1,\,\,\, ~~~\label{eq23a}
\eeq 
\beq X_1= cn (x+x_0|m),\,\,\,
\lambda_1=0,\,\,\, ~~~\label{eq23b} 
\eeq
\beq
 X_2= sn (x+x_0|m),\,\,\,
\lambda_2=m,\,\,\, ~~~\label{eq23c}
\eeq
while other eigenfunctions of Eq. (\ref{eq21}) have much
more complicated forms.
It is worth stressing that the functions in 
(\ref{eq23a},\ref{eq23b},\ref{eq23c}) cannot be
called eigenfunctions of our problem in (\ref{eq10})-(\ref{eq11}),
because they do not satisfy the boundary conditions (\ref{eq11}).
Taking now into account Eq. (\ref{eq11}) we can find the curves of
neutral stability, i.e. the critical relationship between parameters
($w$ and $H$ or $I$ in our case) when the problem
(\ref{eq10})-(\ref{eq11}) has an eigenvalue $\lambda=0$.

In the following we shall examine the implications of the above three
analytic eigenfunctions on the stability of the static solution. We
must bear in mind though that it is not sufficient that one of the
three above eigenvalues is zero or negative, but on top we must
satisfy the boundary conditions. Even in that case we can prove
instability but not the reverse, which can only be done by numerical
solution of the eigenvalue problem. We will see, however that some of
the conclusions will be very useful. We will examine the situation for
each of the three branches separately.

{\bf Ist branch:}
In this case it is not easy to get analytical estimates.
We do not have an extra neutral stability since
$m<1$ and in any case it can be considered as a continuation of branch
II.

{\bf IInd branch:}
In this case the three eigenfunctions of interest are
\beq X_0= cn \left[\sqrt{m}~x|\frac{1}{m}\right],\,\,\,
\lambda_0=m-1~>~0,\,\,\,  \nonumber \eeq
\beq  X_1= dn
\left[\sqrt{m}~x~|\frac{1}{m}\right] =dn
\left[\frac{x}{\sqrt{\bar{m}}}|\bar{m}\right] ,\,\,\,
\lambda_1=0,\,\,\,~~~\label{eq24} \eeq
\beq  X_2= sn
\left[\sqrt{m}x|\frac{1}{m}\right],\,\,\, \lambda_2=m>0.\,\,\,
\nonumber  \eeq

Again, since $m>1$ only $X_1$ is of interest. It is an eigenfunction
of the linearized problem if

\beq sn\left(\frac{w}{\sqrt{\bar{m}}}|\bar{m}\right) = 0  \label{eq27}
\eeq or equivalently if \beq \frac{w}{\sqrt{\bar{m}}} = 2~j~
K(\bar{m}),~~~ j=1,2,\cdots .  ~~~\label{eq28} \eeq
Substituting (\ref{eq28}) into (\ref{BIIa}) we get two families of curves of
neutral stability, where again we can distinguish two cases for even
$(j=2n)$ and odd $(j=2n+1)$ values of $j$.
For even $j$ we have $N_f=2n+1$ and for odd $N_f=2n$ , i.e. odd and
even number of fluxons correspondingly. The respective values of $w$
are given by \beq w=\frac{8}{H}n K\left(\frac{4}{H^2}\right),
~N_f=2n~~~~,\,\, {\rm with}\,\, n=0,1,\cdots \label{eq29} \eeq \beq
w=\frac{2(2n+1)}{\sqrt{1+\frac{H^2}{4}}}
K\left(\frac{1}{1+\frac{H^2}{4}}\right), ~~~N_f=2n+1~~~~,
\,\,{\rm with} \,\, n=0,1,\cdots . \label{eq30} \eeq

{\bf IIIrd branch:} In this case $m>1$  and we can write the
eigenfunctions in the following form
$$ X_0= cn \left[\sqrt{m}(x+x_0)|\frac{1}{m}\right],\,\,\,
\lambda_0=m-1 >0,\,\,\,
$$ 
\beq X_1= dn
\left[\sqrt{m}(x+x_0)|\frac{1}{m}\right]
=\frac{\sqrt{1-\frac{1}{m}}}{dn \left[\sqrt{m}~x|\frac{1}{m}\right]}
=\frac{\sqrt{1-\bar{m}}}{dn
\left[\frac{x}{\sqrt{\bar{m}}}|\bar{m}\right]} ,\,\,\,
\lambda_1=0,\,\,\,~~~\label{eq31} 
\eeq 

$$ X_2= sn
\left[\sqrt{m}(x+x_0)|\frac{1}{m}\right],\,\,\, \lambda_2=m>0 \,\,\,
$$

In (\ref{eq31}) we used a standard transformation of elliptic
functions so that their modulus is less than unity, and then
substituted $x_0=\sqrt{\frac{1}{m}} K(\frac{1}{m})$. We also used the
notation $\bar{m}=\frac{1}{m}$. Since $\lambda_0$ and $\lambda_2$ are
always positive we only need to consider $X_1$. For $X_1$ in
(\ref{eq31}) to be an eigenfunction of (\ref{eq10})-(\ref{eq11}) we
must have again the condition (\ref{eq27}) or the equivalent
(\ref{eq28}).
Here we can distinguish two cases for even
$(j=2n)$ and odd $(j=2n+1)$ values of $j$.
It can easily be verified again
that for even $j$ we have $N_f=2n$ and for odd $N_f=2n+1$ , i.e. even
and odd number of fluxons correspondingly.

Substituting Eq. (\ref{eq28}) into Eq. (\ref{BIIIa}) we obtain two
families of curves of neutral stability in $H$, with the values of $w$
given for $n=1,2,\cdots$ by \beq  w=\frac{4n}{\sqrt{1+\frac{H^2}{4}}}
K\left(\frac{1}{1+\frac{H^2}{4}}\right), ~~~N=2n-1~~~~\label{eq32}
\eeq and \beq w=\frac{4}{H}(2n-1) K\left(\frac{4}{H^2}\right),
~N=2n.~~~~\label{eq33} \eeq Eq. (\ref{eq32}) is valid for any $H\ge
0 $ but (\ref{eq33}) only for $H\ge 2$. As we will see in the section of the
numerical evaluation of stability the above families of curves will in
fact compare very well giving the boundaries of stability.

\subsection{ Numerical stability results}

\begin{figure}
  \centerline{
     \psfig{figure=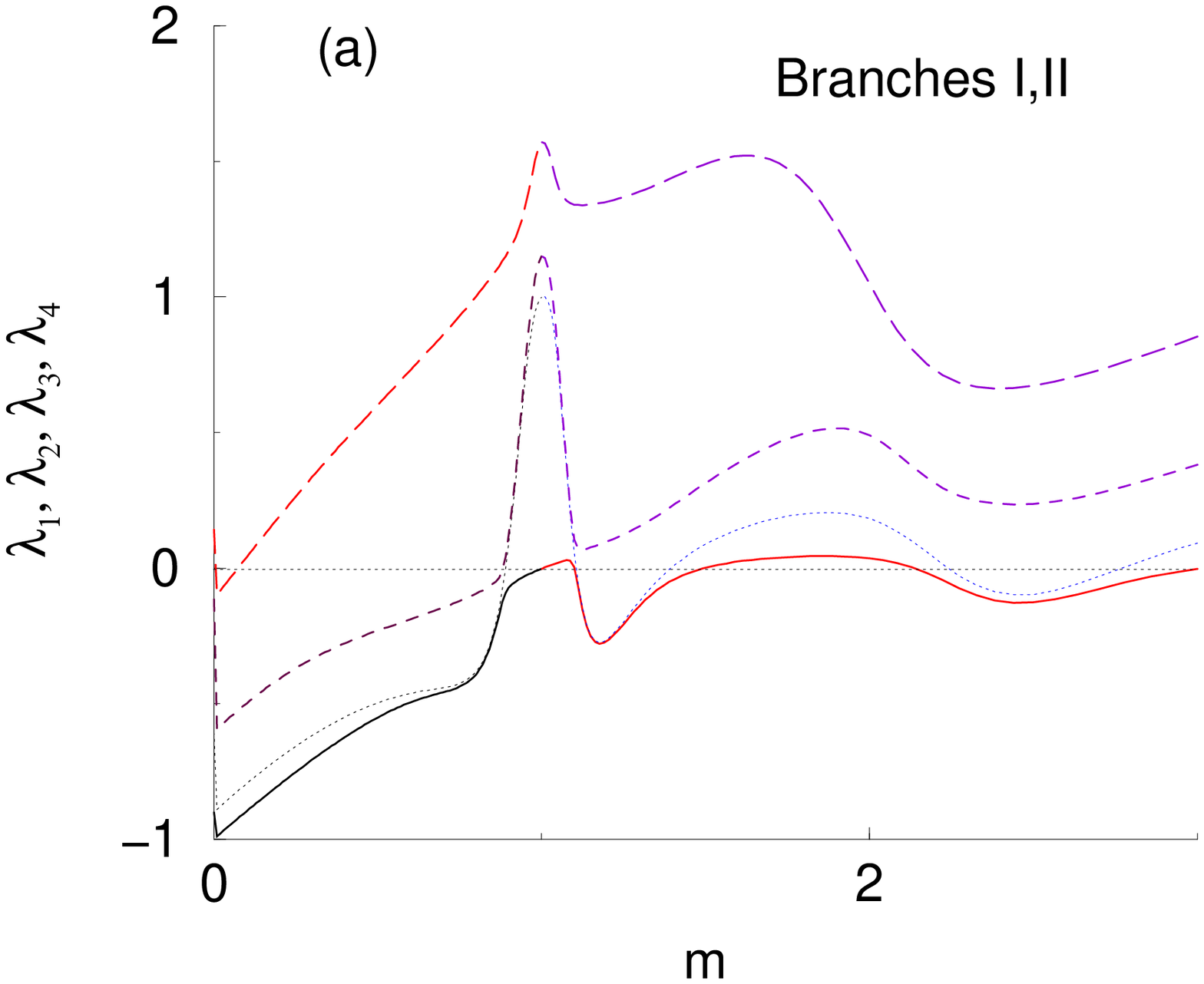,height=2.85in,width=3.0in}
             }
  \centerline{
     \psfig{figure=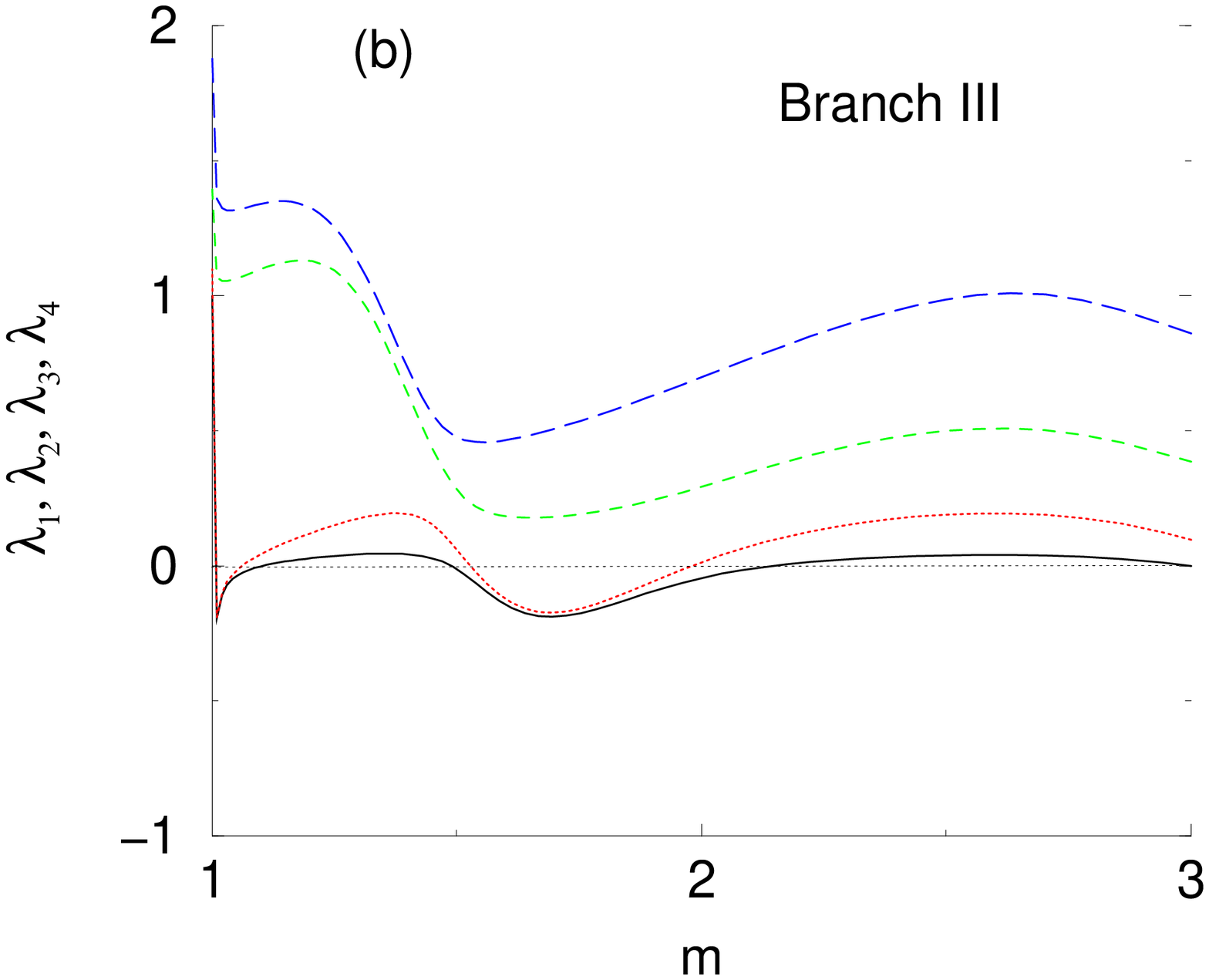,height=2.85in,width=3.0in}
             }
  \caption{Plot of four lowest eigenvalues as a function of $m$
     for (a) branches $I$ and $II$ and (b) branch $III$.}
  \label{fig:12}
\end{figure}

To check the stability of the solutions discussed and examine the
validity of the analytical stability results we have calculated the
eigenvalue spectrum for small oscillations around the solution
$\Phi_0(x)$. In Fig.~\ref{fig:12} we plot the four lowest eigenvalues of Eq.
(\ref{eq10}) for $w=10$ and $I=0$ as a function of the parameter $m$ for the
$I$ and $II$  (Fig.~\ref{fig:12}a) and the $III$ branch (Fig.~\ref{fig:12}b)
correspondingly. Stability requires all eigenvalues to be positive
while an increasing number of negative eigenvalues denotes a higher
degree of instability.

\begin{figure}
  \centerline{
     \psfig{figure=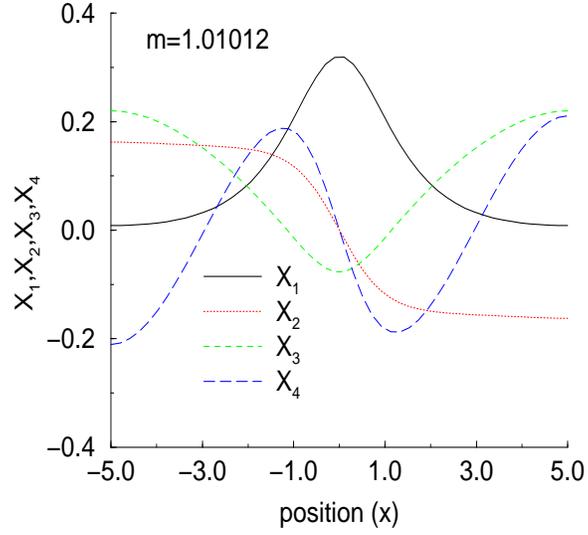,height=2.85in,width=3.0in}
             }
  \caption{Plot of four lowest eigenmodes for $m=1.01012$ of branch 
$II$.}
  \label{fig:13}
\end{figure}

The regions in $m$ with all positive eigenvalues correspond to the
stable fluxon solutions and are separated by regions in $m$ with
unstable solutions. Often  stable solutions correspond (in the plot of
$N_f$ with $H$) to the branches with positive slope, line $ec$, etc in
branch $II$ of Fig.~\ref{fig:8}  or $qp$, etc in branch $III$. This is not
the case though of small $w$ (see Fig.~\ref{fig:7}) or strong magnetic
fields. All the solutions in branch $I$ ($m<1$) are unstable, with
several eigenvalues being negative. In all cases as expected the
lowest mode has no nodes and is symmetric.  It can have however
several lobes, reflecting the number of fluxons that show in the
unperturbed solution $\Phi_0(x)$. Also when a higher mode eigenvalue
vanishes the lowest mode reflects this and reforms by creating more
lobes, and this effect can be strong when eigenvalues cross each other. In
comparing Figs.~\ref{fig:12}a and b we see that the regions of stable
solutions in $m$ for branch $II$ are regions of unstable solutions for
branch $III$, while the bounding values of $m$ are the same in both
cases.  This is consistent with the analytic expressions (26) and
(28). Of course they correspond to different solutions due to
different $x_0$.  In branches $II$ and $III$ there are at most two
negative eigenvalues, while in branch $I$ [$m<1$ in (\ref{eq23c})] there is a
region with four negative eigenvalues.  The analytic formulas for
stability are entirely consistent for the points in $m$ where
instability sets in.  In fact in Fig.~\ref{fig:13} we show the four lowest
eigenvectors for $m=1.01012$ where $\lambda_1=0$. The corresponding
eigenvector is fitted with equation $(25)$ and as expected for
$m\approx 1$ the $dn$ function behaves like a ${\rm sech}(\sqrt{m}x)$.  The
same is true for the branch $III$, where the lowest mode can be fitted
well with $(31)$.

\begin{figure}
  \centerline{
     \psfig{figure=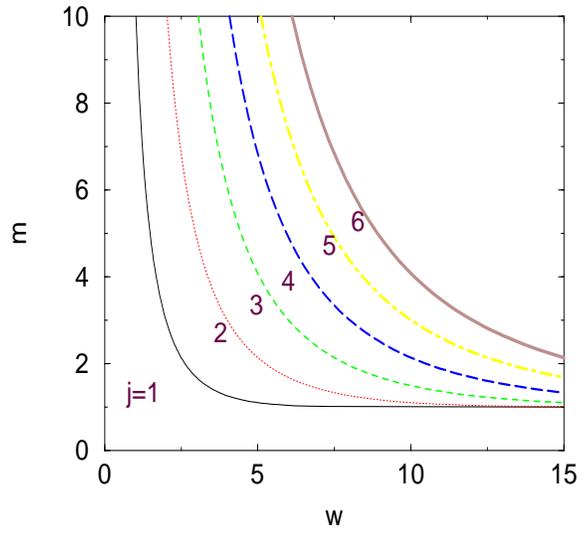,height=2.85in,width=3.0in}
             }
  \caption{Plot of the neutral stability lines
     of integer $N_f$ in an $m$ vs $w$ diagram.}
  \label{fig:14}
\end{figure}

\begin{figure}
  \centerline{
     \psfig{figure=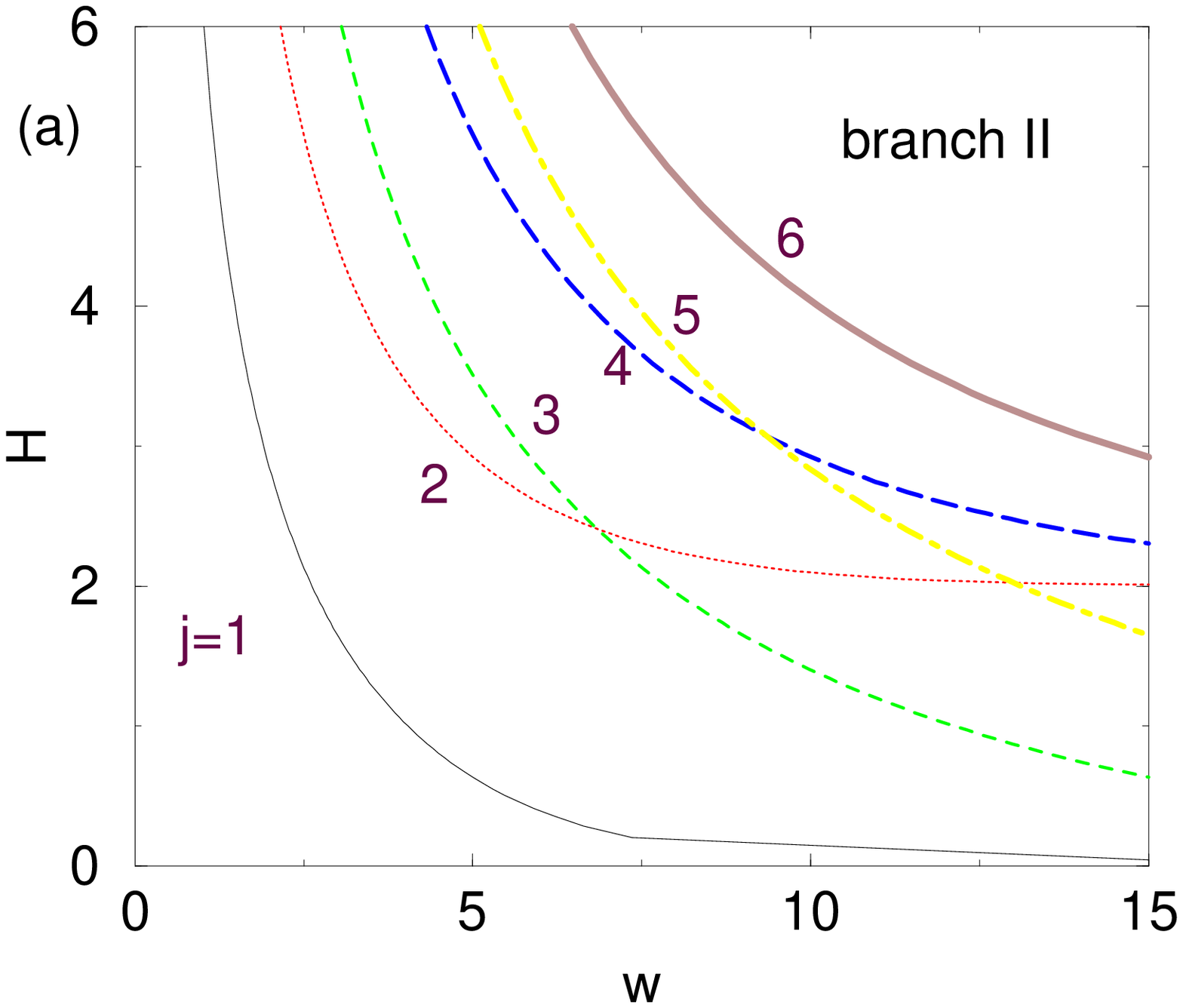,height=2.85in,width=3.0in}
             }
  \centerline{
     \psfig{figure=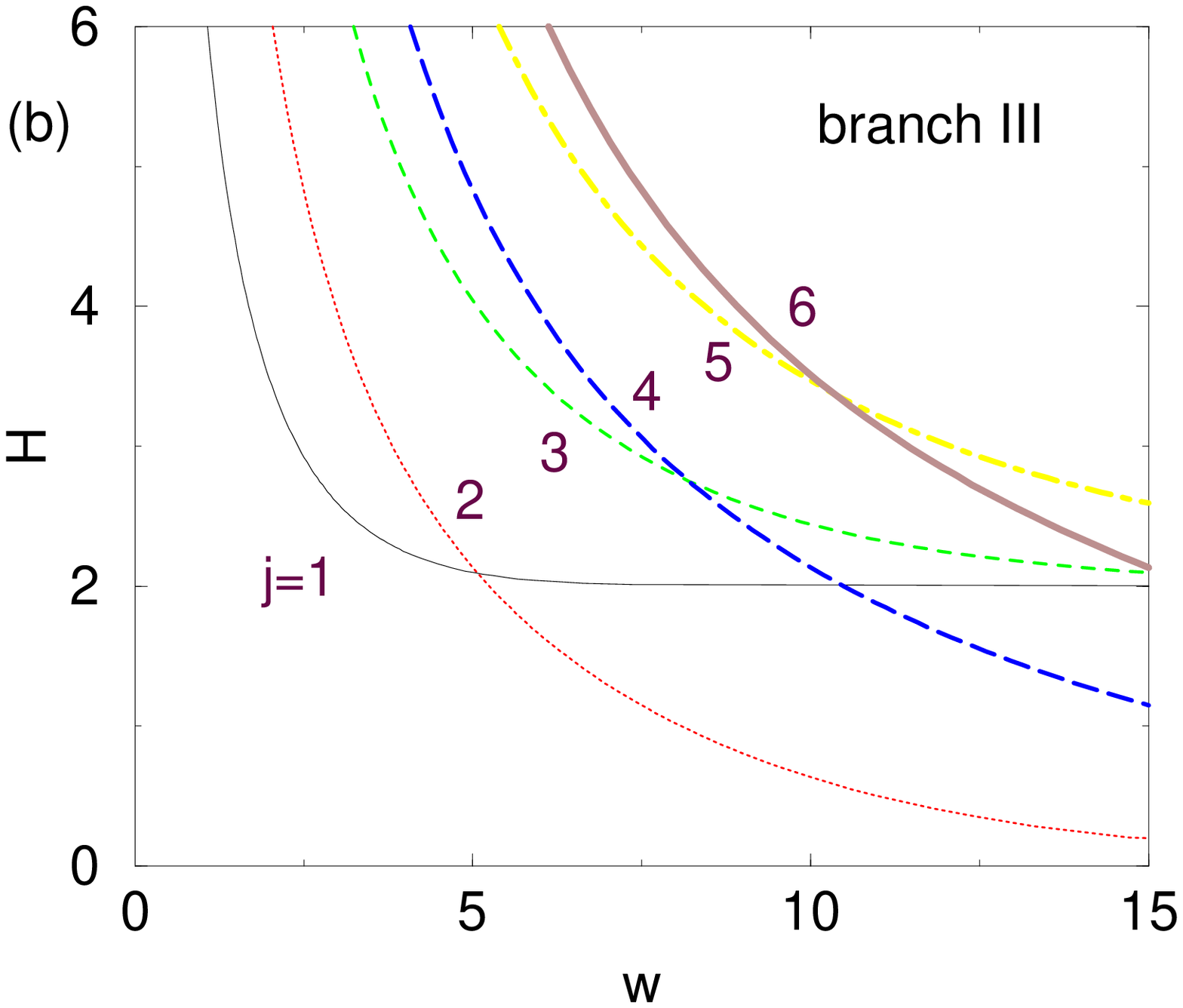,height=2.85in,width=3.0in}
             }
  \caption{ Same as Fig. 14
     in an $H$ vs $w$ plot. (a) branch I and II and (b) branch III.}
  \label{fig:15}
\end{figure}

In Fig.~\ref{fig:14} we plot in the $m$ vs $w$ diagram the lines from conditions
(\ref{eq28}), so that each line corresponds to solutions with an
integer fluxon number. Thus the range of $m$ values between two lines
for a given $w$ corresponds to the $(j,j+1)$ branch of both II and III
cases.  In Fig.~\ref{fig:15} we plot the same information but in an $H$ vs $w$
plot for II (Fig.~\ref{fig:15}a) and III (Fig.~\ref{fig:15}b). Due to the oscillatory
relation of $H$ and $m$ the extrema of the magnetic field at $I=0$ for
each branch do not exactly coincide with the stability boundary lines.
This means that the branch ends on these  lines but in the
intermediate it might reach $H$ values slightly outside this range. If
viewed as a function of $w$ for a fixed $m$ (or $\bar{m}$) then the
corresponding magnetic field varies periodically, with a period in $w$
equal to $4\sqrt{\bar{m}} K(\bar{m})$, i.e. for each $m$ it covers the
$w-$values between every other curve ($\Delta j=2$). We also notice
from Fig.~\ref{fig:15} that the width in $H$ of the (0,1) branch is almost
independent of $w$ for large $w$, while for $w\rightarrow \infty$ many
branches tend to coalesce in the same interval of $H$ for both II and
III. One can understand this case  by using the pendulum analogy
and realize that for large $w$ the solutions correspond to
trajectories that pass near the separatrix points. This can also be
seen from (\ref{BIIIa}). In order to satisfy the boundary condition for
$w\rightarrow \infty$, the important values of $\bar{m}$ are quite
close to $\bar{m}\approx 1$, since in that case
$K(\bar{m})\rightarrow \infty$. On the other hand for high magnetic
fields $H\rightarrow \infty$ the corresponding values of $\bar{m}$ are
near zero.

\section{ Experimental Relevance }

\begin{figure}
  \centerline{
     \psfig{figure=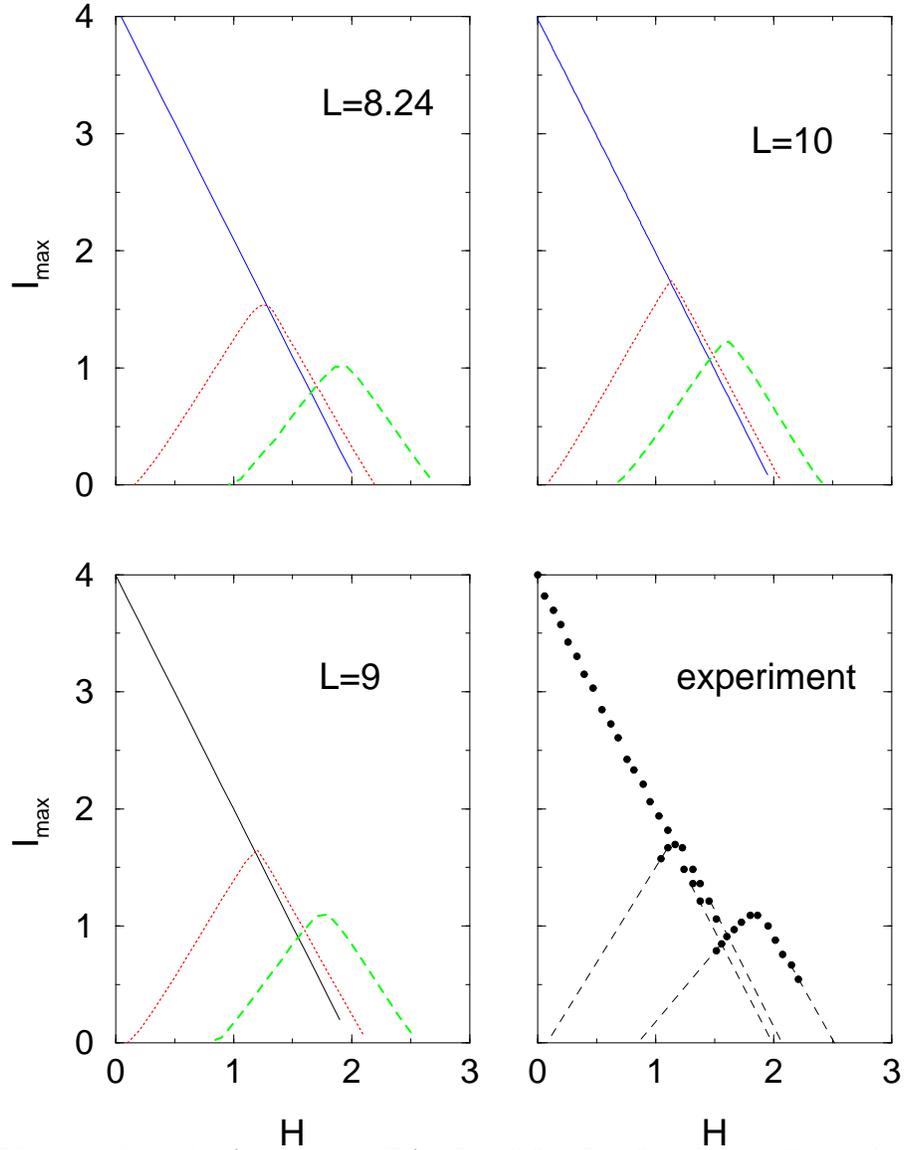,height=6.0in}
             }
\caption{Numerical results for $I_{max}$ vs $H$ for $L=8.24$, 
$L=9$ and $L=10$
and the experimental data from [1]. In the experimental
data the dots are the measured points and the dashed lines
should only be considered as a guide for the eye. }
\label{fig:new}
\end{figure}

In this section we relate some of our theoretical and numerical results
to the experimental techniques and data. Since the behavior of
the maximum tunnel current is of importance for junction
characterization, we start by plotting in Fig.~\ref{fig:new} the
numerically estimated (using the iteration procedure described in
section 4.2) values of $I_{max}$ for three different lengths
(8.24, 9 and 10) and the associated experimental data extracted
from figure 5.10 in \cite{Barone}.
The best fitting seems to be for $L=9.0$, i.e. slightly different
from $L=8.24$, as was determined by analysis of the experimental
data. The discrepancy is due to the fact that the critical current
density is not homogeneous in the experimental sample and the
analysis used in the experiment it would be valid for a larger
length junction. An exact knowledge of the inhomogeneity can give
a more accurate profile but this is not the purpose of this paper.

Besides the relatively good agreement of the maximum current
$I_{max}$ value for each $H$, that verifies our numerical and
theoretical analysis we can also make the following observations:
The experimental data for $I_{max}$ seems to try to follow the
stable branches. After crossing of $I_{max}$ lines it seems to
approach some ''bifurcation" points where it becomes easy to fall
in another branch, but a careful experimental quasistatic scanning
of the magnetic field and current ( as was done in the numerical
simulations used to obtain the displayed data), will enable to
successfully trace the whole stable branch experimentally. Then
one can pass over these 'fuzzy' bifurcation points and be able to
select the appropriate continuation branch. To work within a given
branch with low $I_{max}$ is of interest for low energy (or
current) devices. The quasistatic scanning will also help to
elucidate the physical nature and the practical consequences of
such bifurcation points.

There is a close relation between the experimental and the
computer simulation methodologies for determining the whole
$I_{max}$ line of a device. One of these methodologies is
described (as a numerical scheme) in \cite{Caputo,Gaididei}. It is
based on the failure of the convergence of the associated
iteration method when the bias current exceeds the maximum valued.
We should mark that in this case one has to solve a large set of
PDE problems associated with continuation points on the $I_{max}$
line through fine tuning of $I$ and $H$ along the boundary line.
This requires significant amounts of computer power since its is
prone to the effect of hysterisis. A similar approach is used in
experiments. In this case tracing $I_{max}$ corresponds in
configuring the device on the border line of the branch.
Specifically slightly higher bias currents switch the device from
the pair tunneling to quasiparticle mode. A similar approach (used
here) is to configure the device so it corresponds to a point on
the H-axis (I=0) and slightly increase the current until the above
mode switch will be detected. It is possible but not easy to
realize such "initial configuration". For both cases one has to
introduce some initial flux configuration.

\begin{figure}

  \centerline{
     \psfig{figure=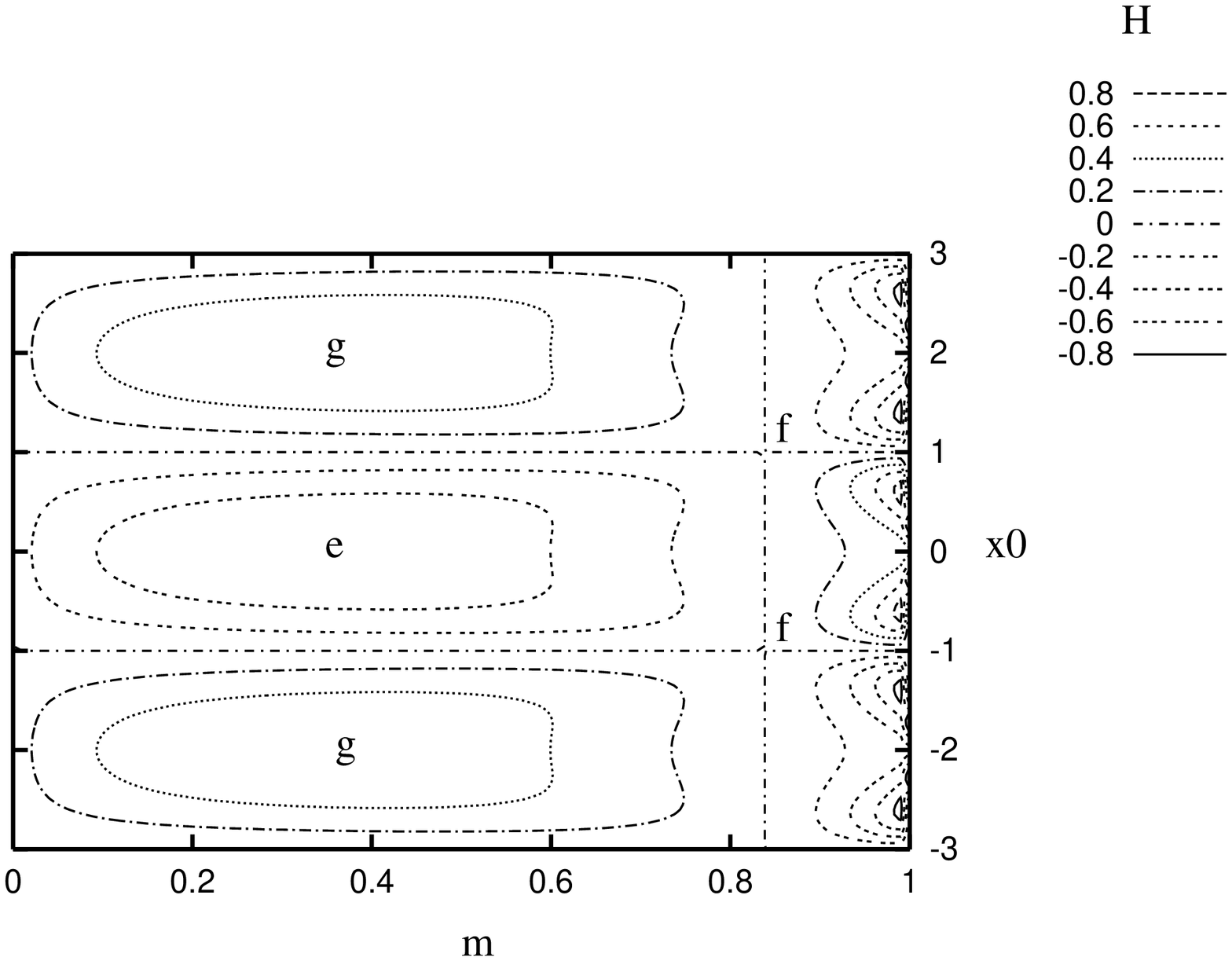,height=2.85in,width=2.5in,angle=-90}
     \psfig{figure=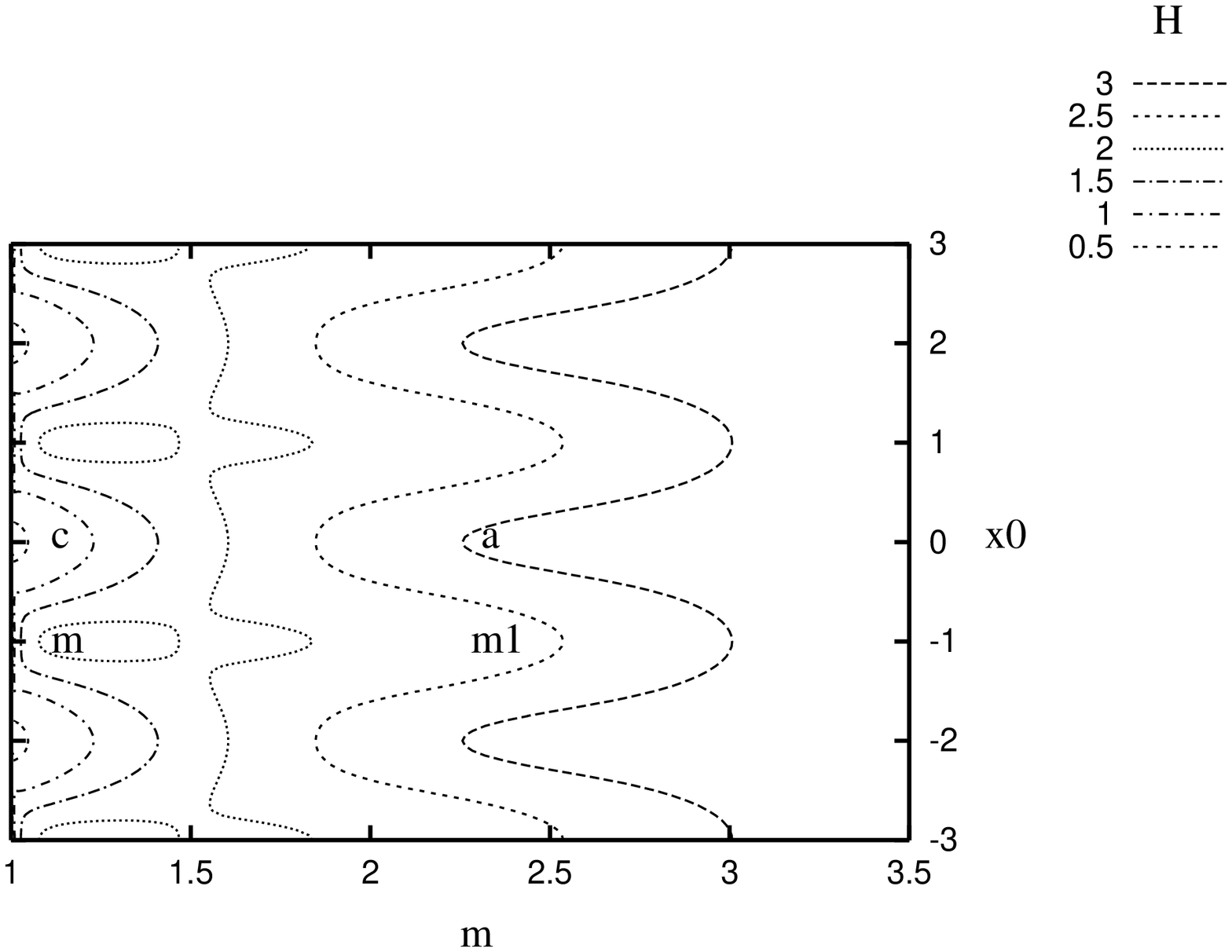,height=2.85in,width=2.5in,angle=-90}
     }
      \centerline{
     \psfig{figure=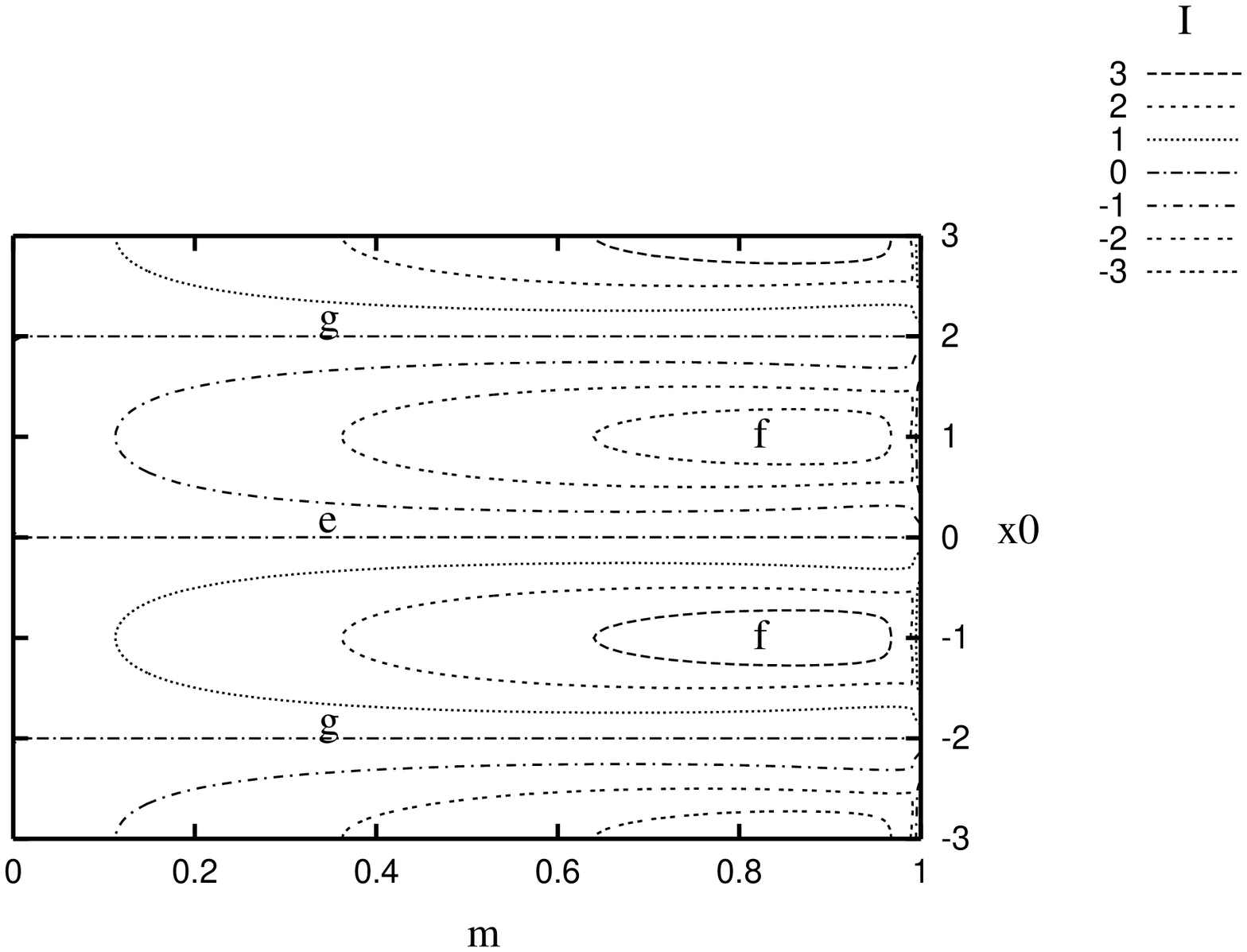,height=2.85in,width=2.5in,angle=-90}
     \psfig{figure=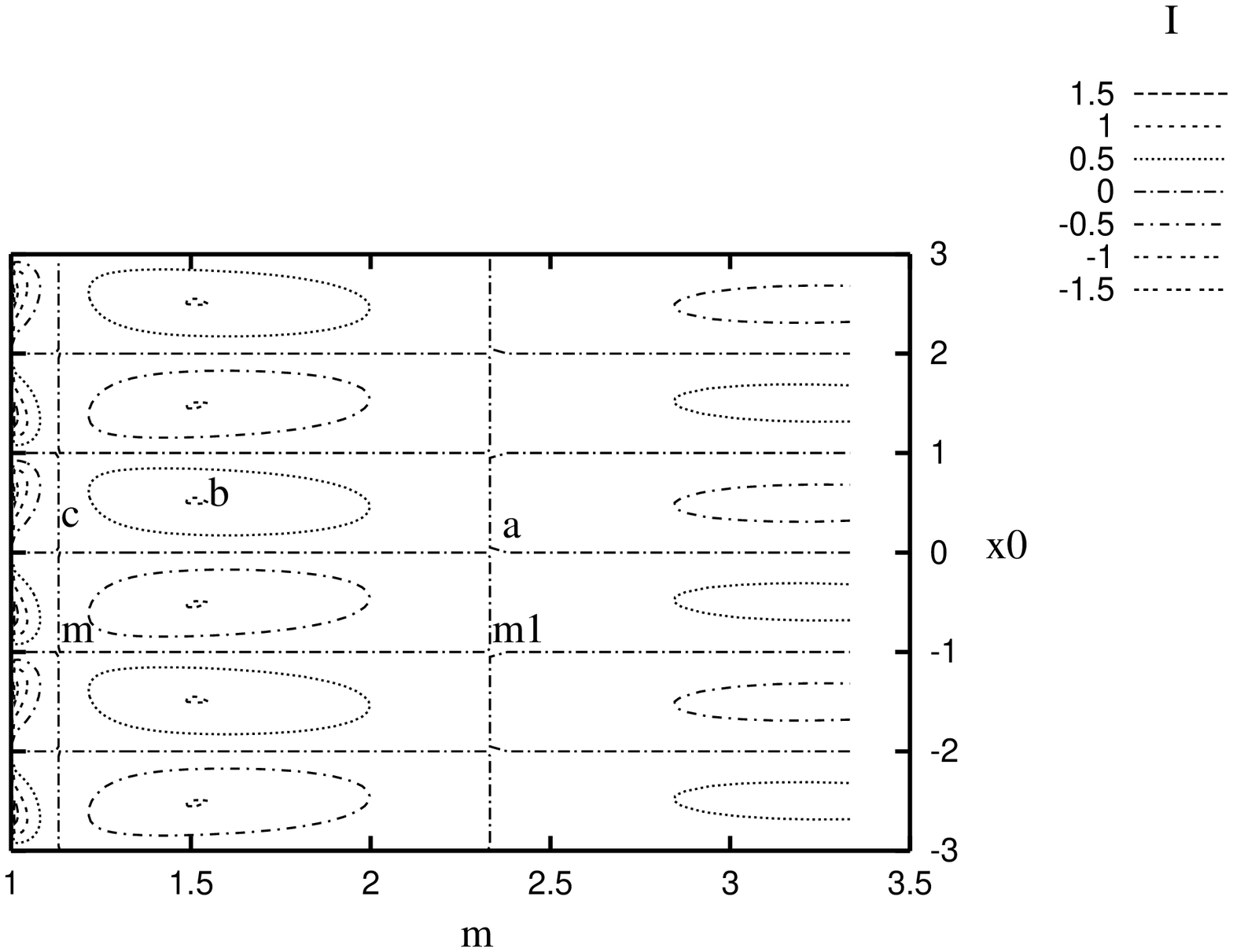,height=2.85in,width=2.5in,angle=-90}
             }
  \caption{Constant $H$ and $I$ contours in the ($m$, $x_0/K(m)$)
plane. }
 \label{fig:17}
\end{figure}

The main difficulty that arises from the hysteretic behavior can be avoided
if one looks at Fig. \ref{fig:17} where we replot Fig. \ref{fig:2}
but with a rescaling of the vertical axis by $K(m)$, so that the
curves $x_0=K(m)$ now become horizontal lines. As we mentioned
earlier the parameters $x_0$ and $m$ define uniquely the fluxon
distribution of the solutions. So the shaded area enclosed by the
curve through the points $a$, $b$, and $c$ (with upper half $I>0$
and lower half $I<0$) is the stable region corresponding to the
($1-2$) branch in the $I_{max}(H)$ diagram (see the corresponding
$a$, $b$ and $c$ points). It is actually separated from the other
stable regions. One can go however from one stable region to the
next, by moving quasistatically along the $x_0=K$ line.

Then it is clear that for an experiment one might select the
starting configuration of his choice and follow an appropriate
path quasistatically to another stable region and to the maximum
current so that the whole procedure is convenient. Note that
keeping $H$ (or $I$) constant corresponds on walking on a
particular contour line.

For the procedure described above where the $H$ is increased or
decreased monotonically the scanning in the $I-H$ plane suffers
from strong hysteretic phenomena that are apparent in both the
computer simulations and the actual experiments. Based on the
analysis presented in this paper an alternative way free of such
phenomena can be very naturally proposed. Specifically, as seen in
Section 4.2 (see in particular Figs.~8a-c) one might consider
searching for the $I_{max}$ on the $I-N_f$ plane where its is
mostly singled valued. The search methodology will remain the same
as before and therefore it can be easily done on computer
simulations by making simple modification on the existing
software. Nevertheless it is not clear how this can be done
experimentally since it requires a manipulation of both the
current and the external field. Another way to keep the flux
constant and this can be done by applying a non-constant magnetic
field that varies trying to keep the distance between the fluxons
constant. It remains to be seen how easily this can be done in
practice.

\section{ Discussion }
In the preceding sections we have presented a theoretical,
numerical and experimental study of the various static solutions
of 1-D Josephson junctions. Our basic approach was the use of
elliptic functions to analytically express the solutions of the
associated sine--Gordon equation. The two parameters involved in
the elliptic functions ($m$ and $x_0$) were properly selected
based on the particular form of the boundary conditions. This let
us obtain useful analytic expressions for these solutions in
particular for the cases of zero magnetic field $H=0$ or zero
current $I=0$. Their importance lays in the fact that one can
easily study their stability by using simple linear perturbations.
This simplicity in the stability analysis let us exploit the role
of the geometric ($w$) and physical parameters ($H, I, N_f$)
involved. A significant outcome of our study is the fact that the
module ($m$) of the elliptic functions is a good characterization
parameter that greatly simplifies the general qualitative and
quantitative pictures of the various solutions. The use of $m$ as
a characterization parameter also leads to more stable, accurate
and efficient numerical algorithms used to study various aspects
of Josephson junctions.

The analysis presented above is particularly useful to understand
the sometimes complicated behavior when we try to follow
numerically the different branches. In fact, this was one of the
motivations behind this work. We are currently building a software
engine to numerically simulate 2 dimensional window Josephson
junctions of various types and configurations. Preliminary
numerical experiments clearly show that although our simulation
engine is build on top of powerful numerical continuation methods
\cite{ag85} using state-of-the-art PDE software
\cite{ellpack,pltmg} very often fails if we do not fully utilize
the results obtained in the current study. Some the most common,
annoying problems encountered (even in the case where no defects
are present) are the following: Considering bifurcation points as
regular point and vice versa, missing bifurcation points, viewing
certain turning or bifurcation points as limiting points and
improper branch switching (e.g. $2\pi$ jumps). The present study
gives several hints to help us drive our simulation engine with no
such problems.

Our original goal was the study of the influence of the critical
current density ($J_c \rightarrow J_c(x)$) inhomogeneities on the
tunneling current $I_{max}$. Two observations, however, made
necessary to study the perfect junction:
\begin{description}
\item[(a)] We noticed that when studying window junctions, even for
zero magnetic field, the maximum current starting with different
initial conditions, was not always the same i.e. at $I_{max}=4.0$.
Several times it stopped at lower values.
\item[(b)] Both
numerical and experimental results show strong hysteresis
phenomena with jumps between different branches when varying the
external magnetic field. Related to this, the question arises
whether there exists a way of analytic continuation between
different branches? Or in more physical terms whether there is a
physical parameter (in the place of $H$) whose smooth variation
shows no hysteresis in $I_{max}$.
\end{description}

With respect to point (a), it is clear now that the second value
belongs to one of the unstable branches we discussed for $w=10$.
If, however, there are defects in the junction the unstable
solutions might also become stable and therefore are of interest
\cite{alexeeva}. Another way to stabilize solutions is by high
frequency fluctuations (of small amplitude) in a way similar to
the Kapitza inverse pendulum problem \cite{landau}. This can also
be achieved by small wavelength spatial variation of the critical
current density \cite{kivshar}. For remark (b) in the undefected
$1-d$ junction one has the advantage that the analytic solution is
known and the choice of $m$ and $x_o$, pin uniquely the proper
solution. Thus one can follow, smoothly the solution if we look at
$I_{max}$ as a function of the magnetic flux. This way we can
avoid hysterisis by choosing a proper initial condition at $I=0$
and increase $I$ to $I_{max}$.

If, however, one uses the magnetic field as an input parameter,
strong hysteretic phenomena are observed, due to the non
uniqueness of the relation between $H$ and $m$ for large $w$. The
multiple solutions (for fixed $x_o$ due to symmetry) correspond to
different fluxon content. This non-uniqueness will disappear for
large $H$ where in fact the junction behaves as if it is a short
one and you recover the diffraction like pattern. Also an increase
of the temperature makes the junction to behave as a short one,
with non-overlapping branches.
 
{\bf Acknowledgments }
 Part of this work was supported by
two PENED grants (No. 2028/1995, 602/1995). Y. G. and J. G. C.
acknowledge the hospitality of the University of Crete. The visit of
J.G.C. was made possible by a grant under the Greek-French
collaboration agreement and the visit of Y. G. by a grant from the
University of Crete.

\begin{table}
\caption{Different types of curves due to various selections of 
$m$ and $x_0$.}
\label{table1}
\begin{tabular}{||l|lr||} \hline
    & $m$                   & $x_0$ (periodic)  \\ \hline
$H=0$   & $0<m<1$           & $\pm(2n+1)K(m) ,n=0,1, ...$   \\ \cline{2-3}
    & $m=m^*, n=0,1, ...n_{max}$    &           \\
    & $\frac{w}{2}=(2n+1)K(m^*)$    & $-\infty<x_0<\infty$  \\
    & if $w>\pi$            &             \\ \hline
I=0 & $0<m<1$       & $x_0=\pm2nK(m)$, n=0,1,... \\ \cline{2-3}
    & $1<m<\infty$      &
$x_0=\pm\frac{1}{\sqrt{m}}2nK(\frac{1}{m}) $  \\ \cline{2-3}
    & $m=m^*<1$         &             \\
    & $\frac{w}{2}=2nK(m^*), w>2\pi$ & $-\infty<x_0<\infty$  \\
    & $n=0,1, ..., n_{max}$     &           \\ \cline{2-3}
    & $m=m^* >1$            &             \\
    & $\frac{w}{2}=\frac{1}{\sqrt{m^*}}nK(\frac{1}{m^*}) $ &
$-\infty<x_0<\infty$    \\
    & $n=0,1, ...$          &             \\ \hline
\end{tabular}
\end{table}

\newpage

\newpage

\end{document}